\documentclass{article}

\usepackage{arxiv}

\usepackage[utf8]{inputenc} 
\usepackage[T1]{fontenc}    
\usepackage{mathptmx}
\usepackage{hyperref}       
\usepackage{url}            
\usepackage{booktabs}       
\usepackage{amsfonts, amssymb}       
\usepackage{nicefrac}       
\usepackage{microtype}      
\usepackage{xcolor}         

\usepackage[left]{lineno}

\usepackage{lipsum}
\usepackage{graphicx, subcaption}
\usepackage{amsmath}
\usepackage[font=small]{caption}
\usepackage{multirow}
\usepackage{longtable}
\usepackage{makecell, makebox}
\usepackage{pifont}
\usepackage{colortbl}
\usepackage{authblk}
\usepackage{enumitem}

\title{Decoding Extrahepatic Targeting of Lipid Nanoparticles with Interpretable Machine Learning
}

\author[1,\textdagger,*]{Asal Mehradfar}
\author[1,\textdagger,*]{Mohammad Shahab Sepehri}
\author[2]{Owen Antholine}
\author[2]{Varun Shankar}
\author[2]{Glen S. Kwon}
\author[1]{Salman Avestimehr}
\author[2,3,*]{Morteza Rasoulianboroujeni}

\affil[1]{Department of Electrical and Computer Engineering, University of Southern California, Los Angeles, CA}
\affil[2]{Pharmaceutical Sciences Division, School of Pharmacy, University of Wisconsin-Madison, Madison, WI}
\affil[3]{\vspace{0.5em}Department of Pharmaceutical Sciences, Gatton College of Pharmacy, East Tennessee State University, Johnson City, TN}
\affil[$\dagger$]{These authors contributed equally to this work.}
\affil[*]{Corresponding authors: \texttt{mehradfa@usc.edu}, \texttt{sepehri@usc.edu}, \texttt{rasoulianbor@etsu.edu}}

\begin{document}

\maketitle

\newcommand{\rc}{\textcolor[rgb]{1,0,0}}
\newcommand{\bc}{\textcolor[rgb]{0,0,1}}
\newcommand{\gc}{\textcolor[rgb]{0,1,0}}
\newcommand{\oc}{\textcolor{orange}}

\begin{abstract}
Lipid nanoparticles (LNPs) have transformed RNA medicine, yet their clinical utility remains constrained by predominant hepatic accumulation after systemic administration. Redirecting LNPs to extrahepatic tissues requires a deeper understanding of how lipid chemistry and formulation composition jointly govern in vivo biodistribution. Here, we develop an interpretable machine learning framework to predict hepatic versus extrahepatic LNP accumulation and identify molecular design rules for extrahepatic RNA delivery. A literature-derived dataset of 476 intravenously administered LNP formulations was curated from 81 studies, integrating formulation composition, lipid chemical structures, and experimentally reported IVIS-based biodistribution profiles. Standardized SMILES representations of ionizable lipids, helper lipids, sterols, PEGylated or polymer-conjugated lipids, additional lipids, and polymer repeat units were converted into RDKit Expert descriptors and combined with formulation-level variables to generate an 808-dimensional feature representation. Logistic regression, random forest, and XGBoost classifiers achieved strong predictive performance on a held-out test set, with ROC–AUC values of 0.839, 0.866, and 0.874, respectively. SHAP-based model interpretation and consensus feature ranking revealed that ionizable-lipid descriptors dominate biodistribution prediction, while formulation composition, particularly ionizable lipid, sterol, and PEGylated/polymer-conjugated lipid fractions, also contributes substantially. Notably, the top 20 consensus features retained nearly all predictive information in tree-based models, indicating that a compact set of molecular determinants is sufficient to classify LNP tropism. The most informative features implicated electrotopological surface properties, charge- and hydrophobicity-weighted surface areas, molecular topology, and amide/alkyl structural motifs as key drivers of extrahepatic accumulation. This study establishes an interpretable, data-driven strategy for decoding LNP biodistribution and provides actionable design principles for engineering next-generation LNPs beyond the liver. Our code and dataset are publicly available at \url{https://github.com/AsalMehradfar/lnp-extrahepatic-targeting/}.

\end{abstract}


\section{Introduction}

RNA-based therapeutics, including messenger RNA (mRNA), small interfering RNA (siRNA), microRNA (miRNA), and antisense oligonucleotides (ASOs), have emerged as powerful tools for modulating gene expression and targeting previously "undruggable" proteins, transcripts, and genes \cite{kim2022rna}. Despite their potential, the clinical application of RNA is limited by its inherent instability, as RNA molecules are highly susceptible to degradation by nucleases in the reticuloendothelial system and within endosomal compartments. Additionally, their inability to passively diffuse across lipid bilayers further hinders their efficacy. Therefore, the development of efficient delivery systems is essential to protect RNA from degradation, enable targeted transport to specific tissues and cell types, and ensure effective cytoplasmic release for optimal therapeutic outcomes \cite{zhu2022rna}.

In 2018, the United States Food and Drug Administration (FDA) approved the first therapeutic small interfering RNA (siRNA), ONPATTRO™, which utilizes lipid nanoparticles (LNPs) as the delivery system for the treatment of hereditary transthyretin-mediated amyloidosis. Similarly, LNP-mediated delivery was successfully employed to transport mRNA encoding the spike protein in COVID-19 vaccines, demonstrating the efficacy of LNPs in RNA drug delivery \cite{yan2022non}. The successful application of LNPs in these RNA-based products underscores their effectiveness as non-viral delivery systems. LNPs are now considered a superior alternative to viral vectors, which face limitations in clinical applications due to high immunogenicity, narrow tissue specificity, costly manufacturing processes, and biosafety concerns \cite{ibba2021advances}. The commercial viability and clinical significance of RNA-LNP formulations have been well established, fueling extensive research into their application for diverse purposes, including cancer immunotherapy, vaccines, and gene-editing therapeutics \cite{hajj2020potent}.

Despite their notable advantages, LNPs face challenges in achieving broader therapeutic applications, primarily due to their tendency to accumulate in hepatocytes following intravenous administration, a process mediated by endogenous transport mechanisms involving lipoprotein receptors and apolipoproteins \cite{lin2024targeting,sato2020different}, leading to rapid clearance from the bloodstream and limiting their use predominantly to liver-related diseases \cite{woitok2020lipid,truong2019lipid}. However, recent preclinical studies have demonstrated that modifications in lipid structures and LNP composition can facilitate the delivery of RNA to extrahepatic organs, such as the lungs and spleen, thereby broadening their therapeutic potential beyond hepatic applications. For instance, Cheng et al. \cite{cheng2020selective} developed Selective Organ Targeting (SORT) technology, which incorporates a supplemental lipid, termed a SORT molecule, into the LNP formulation to achieve tissue-specific mRNA delivery to the lungs or spleen. Additionally, Ni et al. \cite{ni2022piperazine} and Gan et al. \cite{gan2020nanoparticles} engineered novel piperazine-containing lipids and adamantyl-containing phospholipids, respectively, that preferentially deliver mRNA to immune cells in vivo without the need for targeting ligands. Radmand et al. \cite{radmand2024cationic} demonstrated that incorporating cationic cholesterol with a cationic helper lipid enhances mRNA delivery to the heart and various lung cell types, including stem cell-like populations. Furthermore, Qiu et al. \cite{qiu2022lung} synthesized a new ionizable lipid that alters LNP tropism, favoring mRNA delivery to the lungs. These advancements underscore the potential of next-generation LNP platforms to overcome current limitations in extrahepatic delivery and broaden the scope of RNA-based therapeutics.

While these advancements suggest that it is possible to achieve extrahepatic tropism by altering the chemical identity of LNPs, specifically through modifications in lipid structures and compositions, the precise relationship between chemical identity and biological function remains poorly understood. As a result, the development of new LNP formulations often relies on screening large libraries of candidates to evaluate their in vivo biodistribution in animal models, such as mice \cite{liu2021membrane,liu2021zwitterionic}. Although emerging technologies, including DNA/RNA barcoding coupled with sequencing \cite{radmand2024cationic,radmand2023transcriptional,guimaraes2019ionizable,dahlman2017barcoded}, have facilitated in vivo high-throughput screening, exclusive dependence on experimental approaches presents significant limitations: (i) high costs associated with scaling up throughput and (ii) the immensity of the chemical space, which encompasses billions of potential lipid structures and compositions. This complexity arises from the four key components of LNPs—ionizable lipids, helper lipids, sterols, and PEGylated lipids—each playing a critical role in determining the properties and functions of the LNPs \cite{albertsen2022role}. Given the impracticality of exhaustive experimental screening, particularly for extrahepatic delivery where the likelihood of success is low, the development of new, more efficient strategies for LNP discovery is critically needed.

Machine learning (ML) has emerged as a powerful approach for accelerating materials discovery and rational design by enabling accurate prediction of material properties from compositional and structural data. In the field of LNPs, ML models have recently demonstrated considerable success in correlating the chemical identity of LNPs with their biological activity. These approaches include supervised binary classification models for predicting high versus low transfection efficiency \cite{ding2023machine,li2024accelerating}, multiclass classification models for predicting different levels of transfection performance \cite{moayedpour2024representations}, and deep learning frameworks integrated with combinatorial chemistry for the design and optimization of ionizable lipids \cite{xu2024agile,Xu2026lumi}. More recently, benchmarking studies have systematically compared different molecular representations and ML algorithms for predicting LNP transfection performance, providing valuable guidance for future model development \cite{mehradfar2025lantern,zahed2026machine}.

Despite recent advances, existing ML models for LNP design have important limitations that restrict their translational utility. First, many studies focus primarily on ionizable lipid structure while treating other formulation variables—including helper lipids, sterols, PEGylated lipids, additional lipid components, and overall lipid composition—as fixed. As a result, these models do not fully capture the combinatorial effects of formulation components that collectively determine LNP biological activity. Second, many current models offer limited mechanistic interpretability, making it difficult to extract actionable design principles for developing new lipid structures or optimized LNP formulations. Third, most ML approaches remain focused on predicting in vitro transfection efficiency. Although useful for high-throughput screening, such predictions may have limited translational relevance because the relationship between in vitro transfection, in vivo biodistribution, and therapeutic efficacy remains incompletely understood. 

Emerging data-driven efforts in in vivo LNP biodistribution are primarily focused on high-throughput platforms, such as barcoded nanoparticle libraries. While these approaches generate large-scale datasets, they primarily quantify nanoparticle accumulation or cellular uptake in specific organs or cell populations and do not necessarily measure functional RNA delivery, transfection, or protein expression. In contrast, in vivo imaging-based biodistribution studies provide imaging readouts linked to reporter expression and therefore offer a functional measure of delivery efficiency. However, the lower throughput of imaging-based studies, combined with the difficulty of extracting standardized quantitative data from the literature, has limited their use in predictive modeling. Thus, there remains a critical need for interpretable, data-driven frameworks that leverage functional biodistribution data to predict tissue-specific LNP delivery and elucidate how lipid chemistry and formulation composition jointly govern extrahepatic accumulation, functional transfection, and protein expression. Such approaches are essential for the rational engineering of therapeutically relevant RNA delivery systems and for expanding the clinical potential of LNP-based nanomedicines beyond the liver.

In this study, we present an interpretable, data-driven framework for predicting the functional in vivo biodistribution of LNPs while simultaneously identifying actionable molecular and formulation design principles for the rational development of next-generation LNPs. We developed supervised machine learning models using a curated dataset comprising 476 LNP formulations extracted from the published literature and performed a comprehensive meta-analysis through feature importance analysis and consensus feature ranking. The dataset integrates lipid chemical structures, formulation compositions, and corresponding in vivo biodistribution profiles, enabling the classification of formulations as either hepatic- or extrahepatic-accumulating. By combining predictive modeling with model interpretability, this framework reveals how molecular- and formulation-level features collectively influence tissue tropism, providing mechanistic insights and practical design rules for engineering LNPs with enhanced extrahepatic delivery for RNA therapeutics.


\section{Dataset}\label{sec:dataset}

To investigate the molecular determinants governing hepatic and extrahepatic biodistribution of LNPs, we constructed a literature-derived dataset by systematically curating experimentally validated LNP formulations reported in published in vivo studies. For each formulation, the dataset includes formulation composition, biodistribution outcomes, and the chemical structures of the constituent lipid components. The curated chemical structures were subsequently used to generate molecular descriptors, enabling a unified molecular representation for subsequent computational analysis. The following subsections describe the publication-based data curation process and the construction of the molecular feature representation used throughout this study.

\subsection{Publication Selection and Data Extraction}

The following inclusion criteria were used to select LNP formulations for this study: (i) formulations were administered intravenously; (ii) biodistribution data obtained by in vivo imaging system (IVIS) imaging were available for multiple organs; (iii) formulations did not contain active targeting ligands; and (iv) the complete chemical structures of all lipid components could be extracted from published sources.

A Google Scholar search covering publications from January 2011 to January 2025 was conducted using the keywords “in vivo biodistribution,” “lipid nanoparticle,” “IVIS,” “mRNA,” and “ionizable.” A total of 126 publications were identified, of which 81 satisfied the inclusion criteria and were included in the final analysis. The complete list of selected publications is provided in Table~S1 in the Supplementary Information.

Each study was systematically reviewed to extract the composition of the LNP formulations, including the identities and molar ratios of ionizable lipids, helper lipids, PEGylated lipids or alternative polymer-conjugated lipids, and sterols. For formulations containing an additional lipid component, its identity and molar ratio were also recorded. For polymer-conjugated lipids, whether PEGylated or conjugated to an alternative polymer, the molecular weight and monomer structure of the corresponding polymer were recorded to ensure consistent structural representation.

Experimental biodistribution profiles were used to assign binary classification labels. Formulations exhibiting the highest IVIS signal intensity in the liver were classified as \emph{liver-accumulating}, whereas formulations showing maximal signal intensity in any other organ were classified as \emph{non-liver-accumulating}. This curation process resulted in a dataset comprising 476 distinct LNP formulations, including 307 unique ionizable lipids, 29 helper lipids, 2 sterols, 34 additional lipids, and 24 PEGylated or polymer-conjugated lipids. Of the 476 formulations, 237 were classified as non-liver-accumulating and 239 as liver-accumulating.

\begin{figure}[ht]
	\centering
	\begin{subfigure}[t]{0.49\textwidth}
            \centering
            \includegraphics[width=0.99\textwidth]{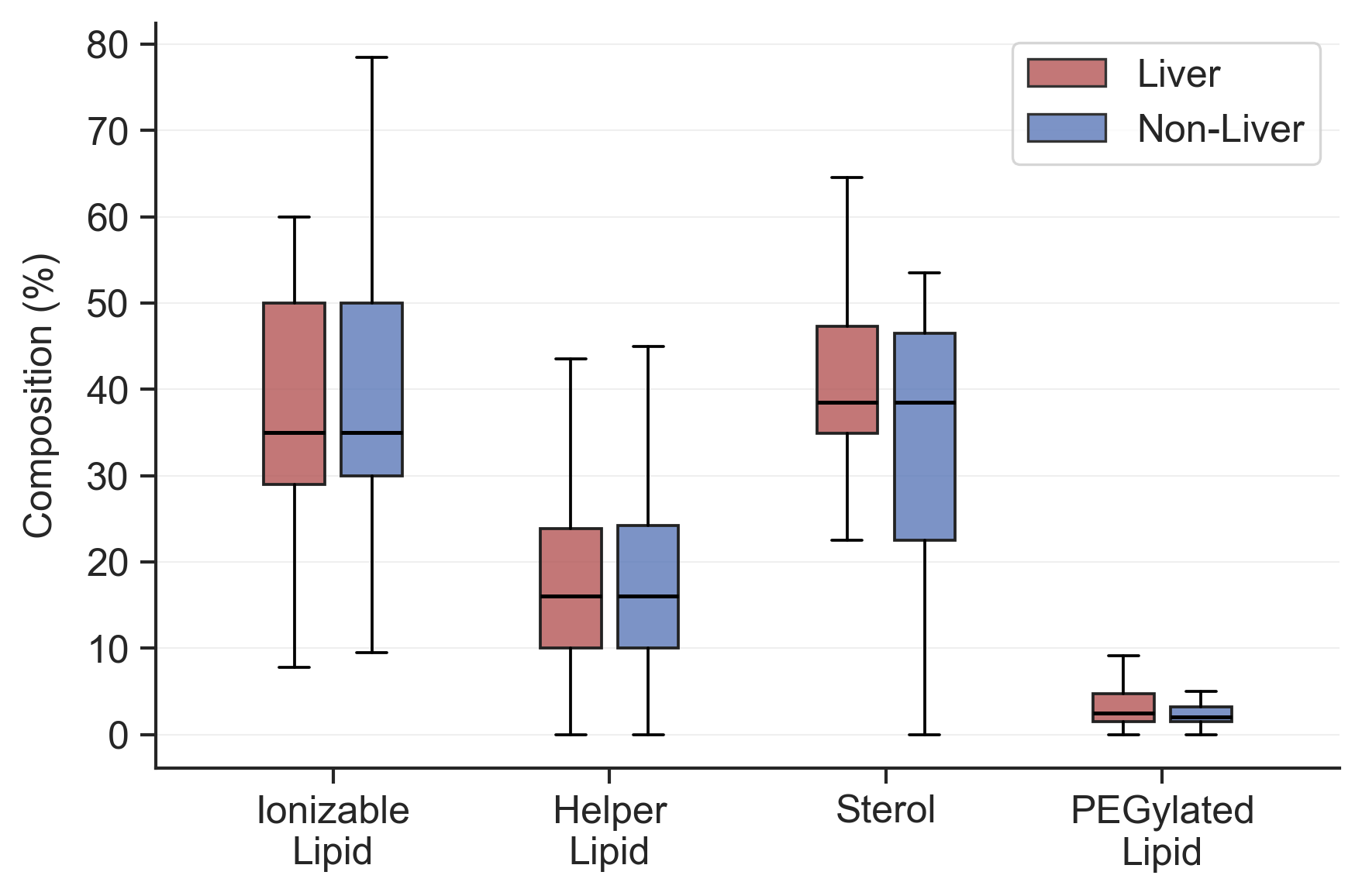}
            \caption{}
        \end{subfigure}
    \hfill
	\begin{subfigure}[t]{0.49\textwidth}
            \centering
            \includegraphics[width=0.99\textwidth]{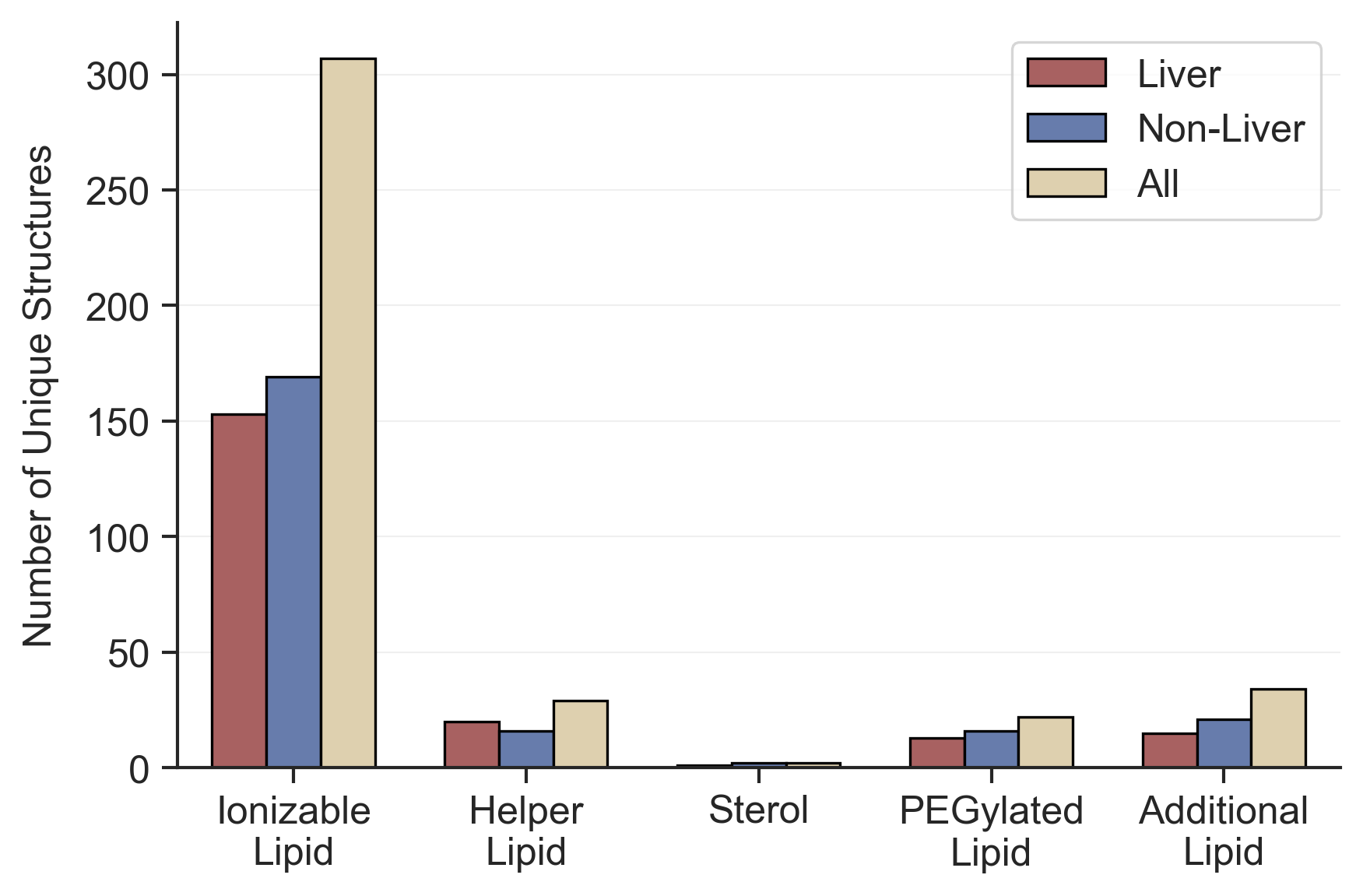}
            \caption{}
        \end{subfigure}
    \caption{\textbf{Overview of the curated LNP dataset.}
    (a) Distribution of lipid composition across liver and non-liver LNP formulations. Box plots show the mole fraction (\%) of the major lipid components, including ionizable lipids, helper lipids, sterols, and PEGylated lipids. The center line represents the median, the box denotes the interquartile range, and whiskers indicate the data spread.
    (b) Chemical diversity of lipid components in the dataset measured by the number of unique molecular structures (unique SMILES) observed for each lipid category. Counts are shown separately for liver and non-liver formulations, together with the total number of unique structures across the dataset.
    }

	\label{fig:data_stats}
\end{figure}

Additional details regarding the distributions of formulation compositions between the two classes, as well as the numbers of unique lipid structures across lipid categories, are provided in Figure~\ref{fig:data_stats}.

\subsection{Molecular Descriptor Extraction and Feature Construction}

Canonical SMILES representations were curated for each molecular component present in the LNP formulations, including ionizable lipids, helper lipids, sterols, PEGylated lipids, additional lipids, and the corresponding polymer component. Molecular descriptors were computed from the standardized SMILES representations using the RDKit cheminformatics toolkit~\cite{landrum2013rdkit, rdkitdesc} with the RDKit Expert descriptor set. This descriptor collection encompasses a broad range of physicochemical, constitutional, topological, and electronic properties derived directly from molecular graph representations.

Descriptors were calculated independently for each molecular component to preserve their individual structural characteristics. For formulations lacking one or more molecular components, the corresponding descriptor fields were encoded in a consistent manner to maintain a uniform feature representation across all formulations. Descriptor columns exhibiting no variation across the curated dataset were subsequently removed. Following this preprocessing step, the final molecular descriptor set comprised 802 numeric features, including 158 descriptors derived from ionizable lipid SMILES, 147 from helper lipid SMILES, 102 from sterol SMILES, 147 from PEGylated lipid SMILES, 162 from additional lipid SMILES, and 86 from the corresponding polymer SMILES. For clarity, molecular descriptor names are prefixed according to the LNP component from which they are derived, including ionizable lipid (IL), helper lipid (HL), sterol (SL), PEGylated lipid (PL), and additional lipid (AL). For example, IL-BalabanJ denotes the BalabanJ descriptor computed for the ionizable lipid, whereas PL-VSA EState7 denotes the corresponding descriptor computed for the PEGylated lipid.

\begin{figure}[ht]
	\centering
        \includegraphics[width=0.85\textwidth]{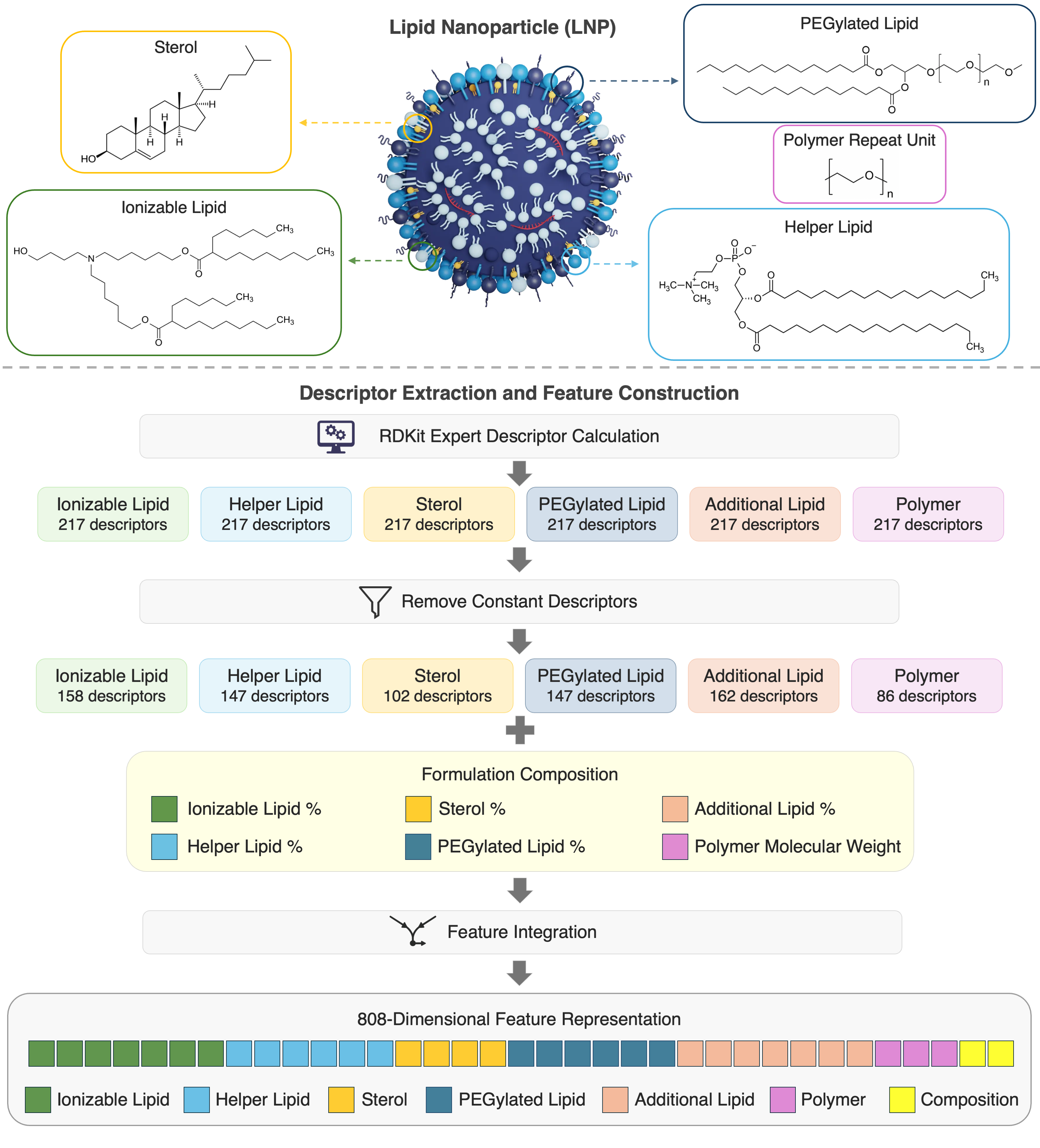}
        \caption{\textbf{Construction of the molecular feature representation.} Representative LNP components, including the ionizable lipid, helper lipid, sterol, PEGylated lipid, and polymer repeat unit, are first represented by their molecular structures. RDKit expert descriptors are calculated independently for each molecular component, followed by the removal of descriptors with constant values across the dataset. The remaining descriptors are combined with formulation composition variables, including the mole percentages of each lipid component and the polymer molecular weight. Finally, all molecular descriptors and formulation variables are concatenated through feature integration to produce the final 808-dimensional feature representation used for downstream machine learning analyses.}
	\label{fig:feature_overview}
\end{figure}

To construct the final feature representation, the filtered molecular descriptors were combined with formulation-level compositional variables, including the mole percentages of ionizable lipids, helper lipids, sterols, PEGylated lipids, and additional lipids, together with the molecular weight of the corresponding polymer. This integration resulted in a comprehensive 808-dimensional feature representation for each LNP formulation. Figure~\ref{fig:feature_overview} schematically illustrates the workflow for constructing the final feature representation used in the subsequent machine learning analyses.

\section{Machine Learning Framework}
This section describes the machine learning framework used to classify LNP formulations and identify molecular determinants associated with extrahepatic targeting. Supervised learning models were trained using the molecular feature representation described in Section~\ref{sec:dataset} and evaluated using standard classification metrics. Model interpretability was subsequently investigated using SHAP analysis, and feature rankings from multiple models were integrated to identify consensus molecular determinants.

\subsection{Machine Learning Models}
Three supervised machine learning models were employed to classify LNP formulations as liver- or non-liver-accumulating: logistic regression (LR) \cite{friedman2010regularization, hastie2009elements}, random forest (RF) \cite{breiman2001random}, and extreme gradient boosting (XGBoost) \cite{chen2016xgboost}. Logistic regression was selected as a regularized linear baseline, whereas random forest and XGBoost were chosen as complementary tree-based ensemble methods capable of capturing complex nonlinear relationships between molecular descriptors, formulation composition, and biodistribution outcomes. Together, these models provide a balance between interpretability and predictive performance while representing diverse machine learning paradigms commonly used for molecular property prediction.

\subsection{Model Training and Evaluation}

Prior to model training, missing feature values were imputed using the median value of each feature, and missing-value indicators were incorporated to preserve information associated with absent molecular components. Logistic regression was implemented within a preprocessing pipeline including median imputation, missing-value indicators, and feature standardization, followed by elastic-net regularization \cite{zou2005regularization}, whereas random forest and XGBoost were trained using median-imputed features with missing-value indicators but without feature standardization. Hyperparameter values were selected empirically based on preliminary experiments and subsequently fixed for all analyses. Logistic regression employed the SAGA solver with elastic-net regularization ($L_1$ ratio = 0.5, regularization strength $C=1.0$), whereas random forest was trained using 500 decision trees with balanced class weights. XGBoost employed 500 boosting iterations with a learning rate of 0.05, a maximum tree depth of 3, subsampling ratio of 0.8, column sampling ratio of 0.8, and an $L_2$ regularization coefficient of 1.0. All models were trained using a fixed random seed (42) to ensure reproducibility.

The curated dataset was divided into training (80\%) and held-out test (20\%) subsets using stratified random sampling to preserve the class distribution. Model development was assessed using five-fold stratified cross-validation on the training set, with the mean and standard deviation of the ROC-AUC reported across folds. Final model performance was evaluated on the held-out test set using ROC-AUC, accuracy, recall, and F1-score. ROC curves were generated to compare the classification performance of the three models. Model calibration was additionally assessed using calibration curves and the Brier score, which quantifies the agreement between predicted probabilities and observed outcomes, with lower values indicating better calibration.

\subsection{Feature Importance Analysis}

Model interpretability was investigated using SHapley Additive exPlanations (SHAP) \cite{lundberg2017unified}. SHAP values were computed for each trained model using the held-out test set to quantify the contribution of individual features to the predicted class. A linear SHAP explainer was employed for logistic regression, whereas tree-based SHAP explainers were used for the random forest and XGBoost models.

Global feature importance was quantified by computing the mean absolute SHAP value of each feature across all held-out test samples. This aggregation produced a single global importance score for each feature, representing its average contribution to model predictions across the held-out test set. Because missing-value indicators were incorporated during data preprocessing, the corresponding SHAP contributions were merged with those of their associated original features to obtain a single importance score for each molecular or compositional feature. These global feature importance scores served as the basis for constructing a consensus feature ranking across the three machine learning models, as described in the following subsection.

\subsection{Consensus Feature Ranking}

The global feature importance scores obtained from the three machine learning models were first normalized within each model to enable meaningful comparison of feature importance across models. For each model, the normalized feature importance of feature $i$ was computed as

\begin{equation}
\tilde{I}_i^{(m)}
=
\frac{I_i^{(m)}}{\sum_{k=1}^{P} I_k^{(m)}},
\end{equation}

where $I_i^{(m)}$ denotes the global feature importance score of feature $i$ obtained from model $m$, computed as the mean absolute SHAP value across the held-out test samples, and $P$ is the total number of features.

Features within each model were subsequently ranked in descending order according to their normalized feature importance scores, with larger normalized importance corresponding to higher feature importance. When multiple features had identical normalized importance values within an individual model, tied features were assigned the average of their corresponding ranks, resulting in one ranked feature list for each model.

The consensus rank of each feature was then calculated as the median of its ranks across the three machine learning models,

\begin{equation}
R_i^{\mathrm{cons}}
=
\operatorname{median}
\left(
R_i^{\mathrm{LR}},
R_i^{\mathrm{RF}},
R_i^{\mathrm{XGB}}
\right),
\end{equation}

where $R_i^{\mathrm{LR}}$, $R_i^{\mathrm{RF}}$, and $R_i^{\mathrm{XGB}}$ denote the ranks assigned to feature $i$ by the logistic regression, random forest, and XGBoost models, respectively. Features were subsequently sorted according to their consensus ranks, with lower consensus ranks indicating features that were consistently identified as more important across different machine learning models.

When multiple features shared the same consensus rank after median aggregation, ties were resolved using the average normalized feature importance across the three models,

\begin{equation}
\bar{I}_i
=
\frac{
\tilde{I}_i^{\mathrm{LR}}
+
\tilde{I}_i^{\mathrm{RF}}
+
\tilde{I}_i^{\mathrm{XGB}}
}{3},
\end{equation}

where features with larger average normalized importance were assigned higher priority. The resulting consensus ranking was used to identify subsets of the highest-ranked features for downstream analyses and reduced-feature model training.


\section{Results and Discussion}

\subsection{Prediction of Extrahepatic Targeting}

\begin{figure}[ht]
	\centering
        \includegraphics[width=0.45\textwidth]{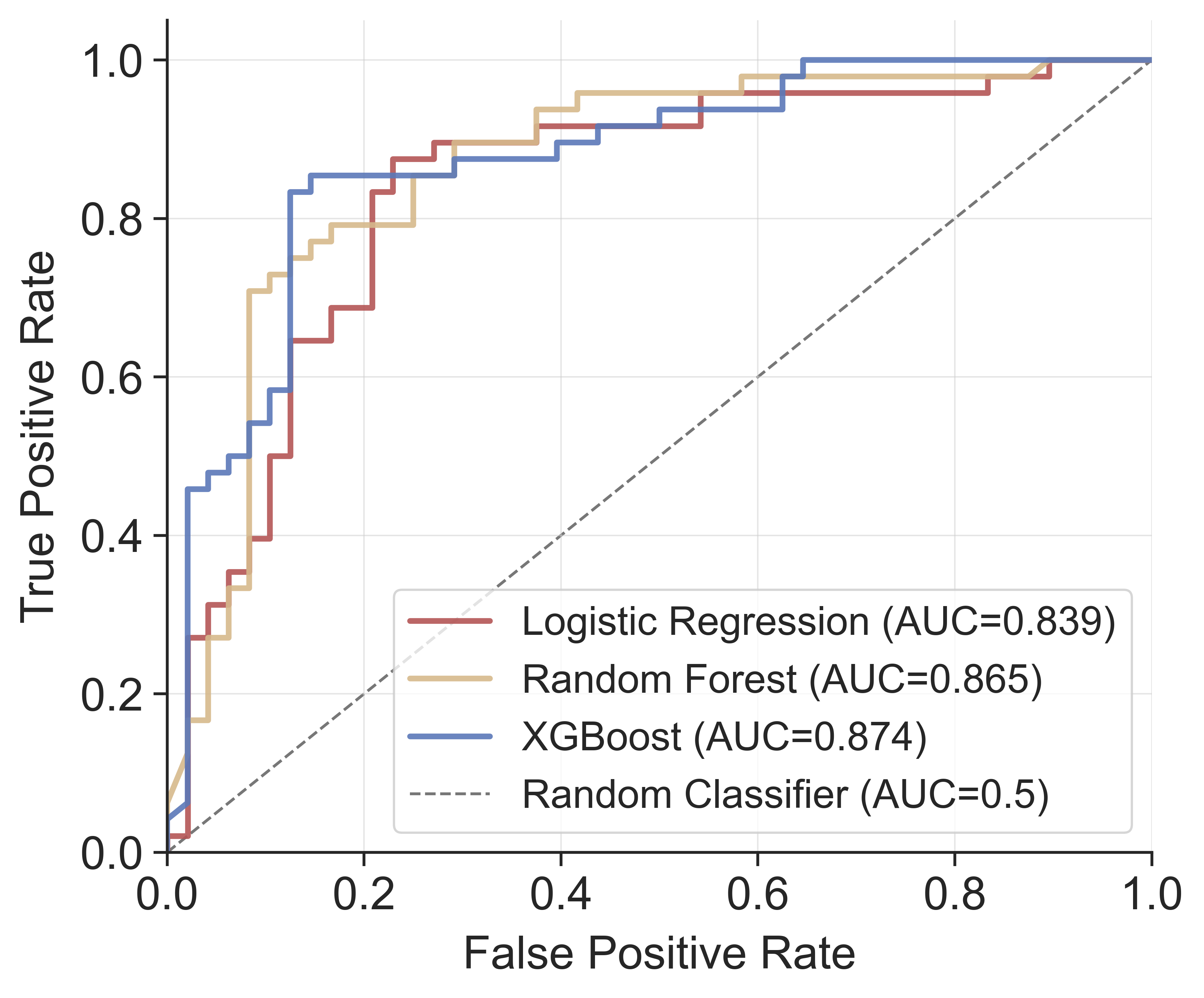}
        \caption{\textbf{Performance comparison of ML models for LNP classification.}
        ROC curves for logistic regression, random forest, and XGBoost models evaluated on the test set. The curves illustrate the trade-off between true positive rate and false positive rate for predicting liver versus non-liver LNP formulations.}
	\label{fig:roc_test}
\end{figure}

The ability of the proposed molecular feature representation to distinguish liver-targeting from extrahepatic-targeting LNP formulations was first evaluated using logistic regression, random forest, and XGBoost classifiers. Receiver operating characteristic (ROC) curves for the three models are presented in Figure~\ref{fig:roc_test}, and the corresponding quantitative performance metrics are summarized in Table~\ref{tab:model_performance}.

All three models achieved strong predictive performance, with test ROC--AUC values exceeding 0.83. Among the evaluated models, XGBoost achieved the highest test ROC--AUC of 0.874, followed closely by random forest (0.866), whereas logistic regression achieved a test ROC--AUC of 0.839. Similar trends were observed for accuracy, recall, and F1-score, with XGBoost consistently providing the strongest overall performance on the held-out test set. Five-fold cross-validation demonstrated comparable performance across training folds, indicating good generalization of the proposed models.

Compared with logistic regression, the superior performance of the tree-based models suggests that the relationship between molecular descriptors and LNP biodistribution is governed by nonlinear interactions that are not fully captured by a linear decision boundary. Nevertheless, the competitive performance achieved by logistic regression indicates that the proposed molecular descriptor representation itself contains substantial predictive information. Calibration analysis further demonstrated good agreement between predicted probabilities and observed outcomes for all three models (Figure~S1). Consistent with the ROC--AUC results, XGBoost achieved the lowest Brier score, followed closely by random forest (Table~S2), indicating reliable probability estimates across the evaluated models.

\begin{table*}[ht]
\centering
\caption{\textbf{Performance of ML models across different feature-set sizes for LNP classification.}
Models were evaluated using the full feature set (\textit{All}) and reduced feature sets containing the top 20 and top 10 features. Cross-validation performance is reported as the mean AUC $\pm$ standard deviation across folds, and held-out test performance is reported using AUC, accuracy, recall, and F1-score. Within each feature-set block, the best-performing values are shown in bold.}
\label{tab:model_performance}
\begin{tabular}{llccccc}
\toprule
\textbf{Feature Set} & \textbf{Model} & \textbf{CV AUC} & \textbf{Test AUC} & \textbf{Accuracy} & \textbf{Recall} & \textbf{F1-score} \\
\midrule

\multirow{3}{*}{All}
& logistic regression & $0.826 \pm 0.054$ & 0.839 & 0.792 & 0.792 & 0.792 \\
& random forest       & $\mathbf{0.882} \pm 0.036$ & 0.866 & 0.802 & 0.792 & 0.800 \\
& XGBoost             & $0.875 \pm 0.034$ & \textbf{0.874} & 0.854 & 0.854 & 0.854 \\
\midrule

\multirow{3}{*}{20}
& logistic regression & $0.759 \pm 0.056$ & 0.724 & 0.615 & 0.583 & 0.602 \\
& random forest       & $\mathbf{0.854} \pm 0.050$ & \textbf{0.866} & 0.781 & 0.812 & 0.788 \\
& XGBoost             & $0.852 \pm 0.048$ & 0.851 & 0.781 & 0.812 & 0.788 \\
\midrule

\multirow{3}{*}{10}
& logistic regression & $0.707 \pm 0.053$ & 0.714 & 0.729 & 0.750 & 0.735 \\
& random forest       & $0.842 \pm 0.054$ & \textbf{0.859} & 0.760 & 0.792 & 0.768 \\
& XGBoost             & $\mathbf{0.852} \pm 0.048$ & 0.847 & 0.771 & 0.771 & 0.771 \\
\bottomrule

\end{tabular}%
\end{table*}

\subsection{Consensus Molecular Determinants of Extrahepatic Targeting}

\begin{figure}[ht]
	\centering
	\begin{subfigure}[t]{0.33\textwidth}
            \centering
            \includegraphics[width=0.99\textwidth]{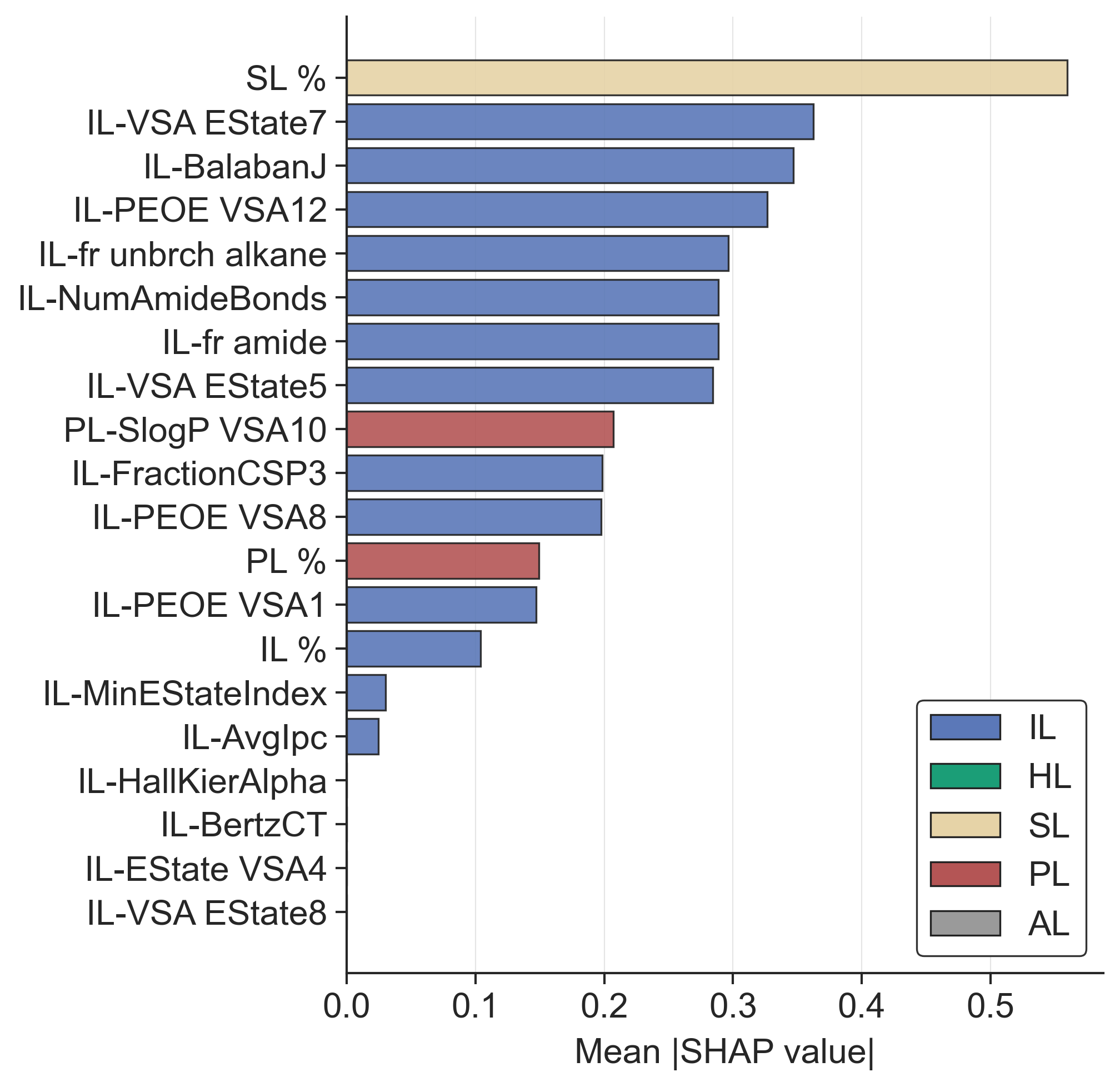}
            \caption{}
        \end{subfigure}
    \hfill
	\begin{subfigure}[t]{0.33\textwidth}
            \centering
            \includegraphics[width=0.99\textwidth]{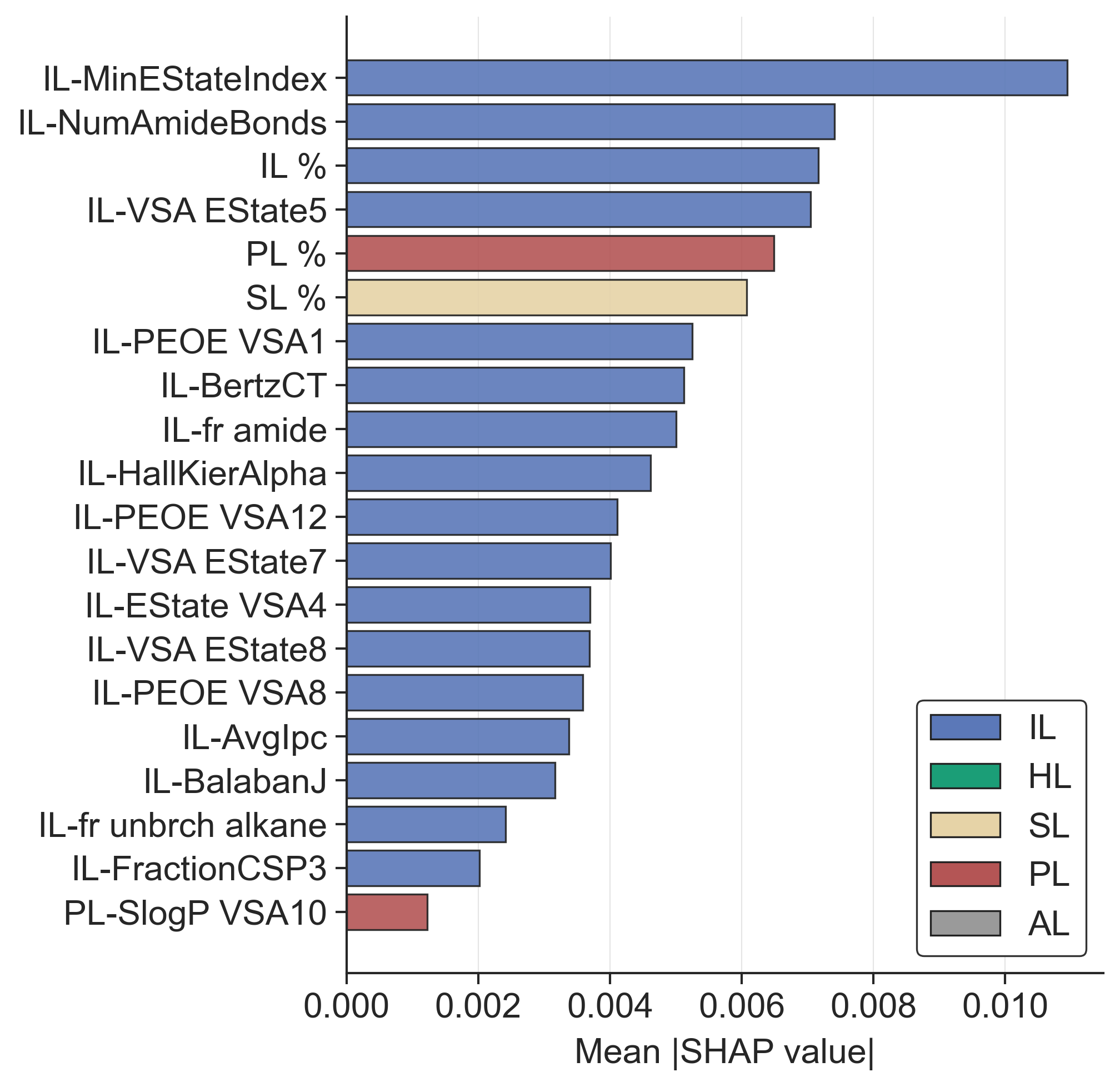}
            \caption{}
        \end{subfigure}
    \hfill
	\begin{subfigure}[t]{0.33\textwidth}
            \centering
            \includegraphics[width=0.99\textwidth]{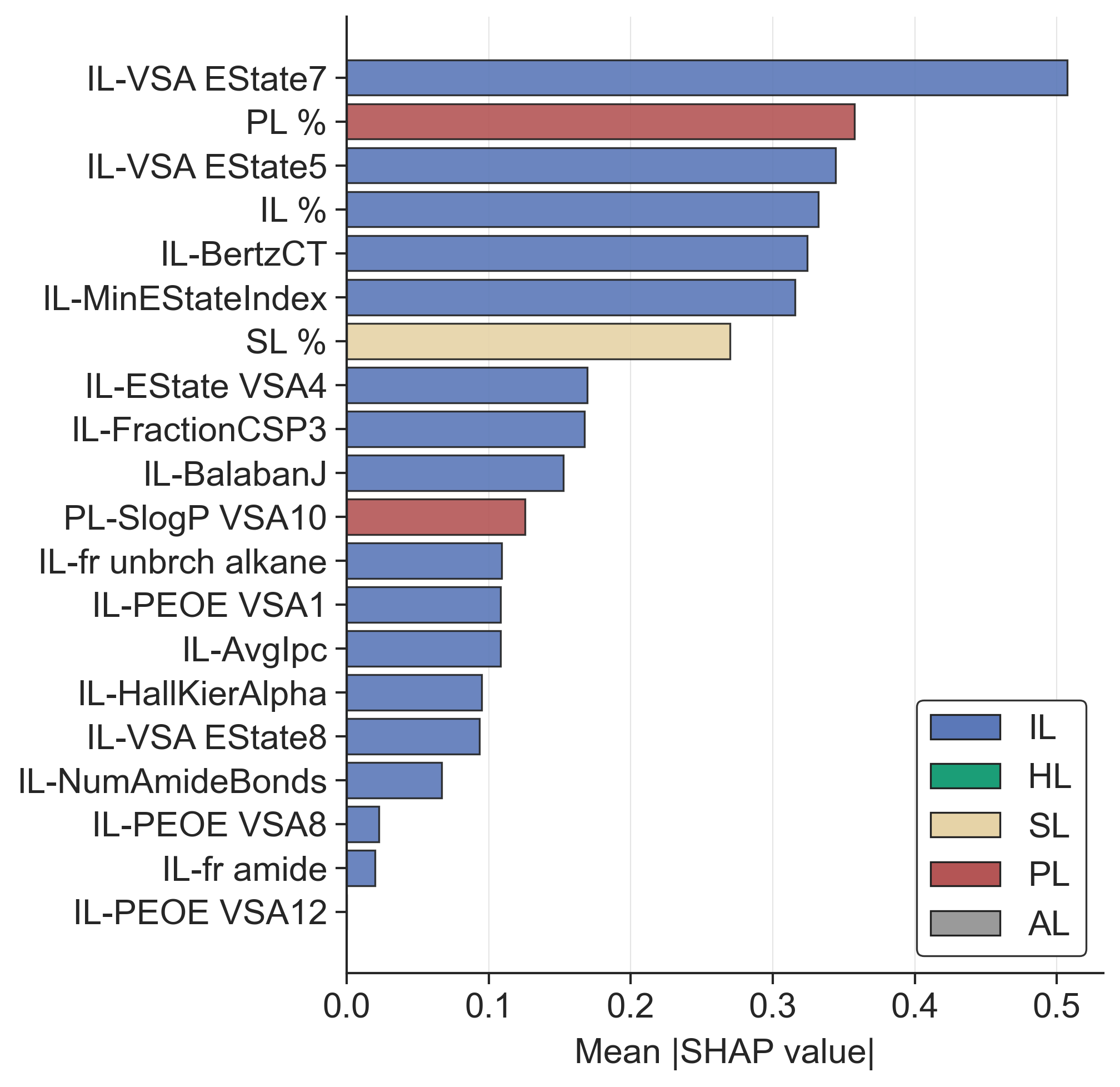}
            \caption{}
        \end{subfigure}
    \caption{\textbf{SHAP-based feature importance analysis of ML models predicting LNP biodistribution.} 
    Mean absolute SHAP values for the top 20 consensus features are shown for (a) logistic regression, (b) random forest, and (c) XGBoost models. Bars are color-coded according to feature categories. Feature names are prefixed by the lipid component from which the descriptor is derived (e.g., IL--), where the suffix denotes the corresponding molecular descriptor or formulation variable. Features are ranked independently within each model, and the relative importance of these consensus features across models highlights key predictors of liver-targeting behavior in LNP formulations.}

	\label{fig:shap_bar}
\end{figure}

To identify molecular descriptors that were consistently informative across different machine learning models, a consensus feature ranking was constructed from the SHAP-based importance rankings of logistic regression, random forest, and XGBoost (Figure~\ref{fig:shap_bar}). Despite differences in the underlying learning algorithms, the three models exhibited substantial agreement in their highest-ranked features, indicating that the identified molecular determinants were robust to the choice of machine learning model. The complete consensus ranking of the top 50 features is provided in Table~S3.

Several molecular descriptors were consistently identified among the highest-ranked features across the three models, including IL-VSA EState5, IL-VSA EState7, IL-MinEStateIndex, and IL-BalabanJ. Although the relative ordering of individual descriptors varied between models, these descriptors repeatedly appeared among the most influential predictors, highlighting their importance in distinguishing liver-targeting and extrahepatic-targeting LNP formulations.

A notable observation is that the majority of the highest-ranked descriptors originated from the ionizable lipid (IL) component of the LNP formulation, suggesting that this component plays a dominant role in distinguishing liver-targeting and extrahepatic-targeting formulations. In addition to molecular descriptors, formulation composition also emerged as an important predictor, with the ionizable lipid percentage (IL \%), sterol percentage (SL \%), and PEGylated lipid percentage (PL \%) consistently appearing among the highest-ranked features across the three models. Together, these findings suggest that both the molecular properties of the ionizable lipid and the relative composition of the LNP formulation play important roles in distinguishing liver-targeting and extrahepatic-targeting formulations.

\subsection{Predictive Power of Consensus Molecular Determinants}

To assess whether the consensus molecular determinants identified in the previous subsection captured the majority of the predictive information, model performance was re-evaluated using reduced feature sets derived from the consensus feature ranking. Specifically, the highest-ranked 20 and 10 consensus features were selected, and the three machine learning models were retrained using only these descriptors. Their predictive performance was subsequently compared with that of models trained using the complete molecular descriptor set (Table~\ref{tab:model_performance}). Performance results for additional feature subsets are provided in Table~S4. The influence of the number of selected features on model performance is further illustrated in Figure~S2.

Reducing the feature set from 808 descriptors to the top 20 consensus-ranked features resulted in only a modest decrease in predictive performance for the tree-based models. Notably, random forest achieved the same test ROC--AUC (0.866) using only the top 20 features as with the complete feature set, while XGBoost exhibited only a slight reduction in test ROC--AUC from 0.874 to 0.851. Although logistic regression showed a larger decrease in predictive performance, the tree-based models consistently maintained superior classification accuracy. These results indicate that the highest-ranked consensus molecular determinants retained nearly all of the predictive information contained in the complete molecular descriptor representation.

Further reducing the feature set to the top 10 consensus-ranked features resulted in a moderate decline in predictive performance for all three models, although random forest and XGBoost continued to outperform logistic regression. As shown in Figure~S2, predictive performance improved substantially as additional consensus-ranked features were incorporated and approached a plateau after approximately 20 features. Collectively, these findings demonstrate that a relatively small subset of consensus molecular determinants is sufficient to accurately distinguish liver-targeting and extrahepatic-targeting LNP formulations while substantially reducing the dimensionality of the molecular descriptor space.

\subsection{Predictive Contributions of Consensus Molecular Determinants}

\begin{figure}[ht]
	\centering
	\begin{subfigure}[t]{0.33\textwidth}
            \centering
            \includegraphics[width=0.99\textwidth]{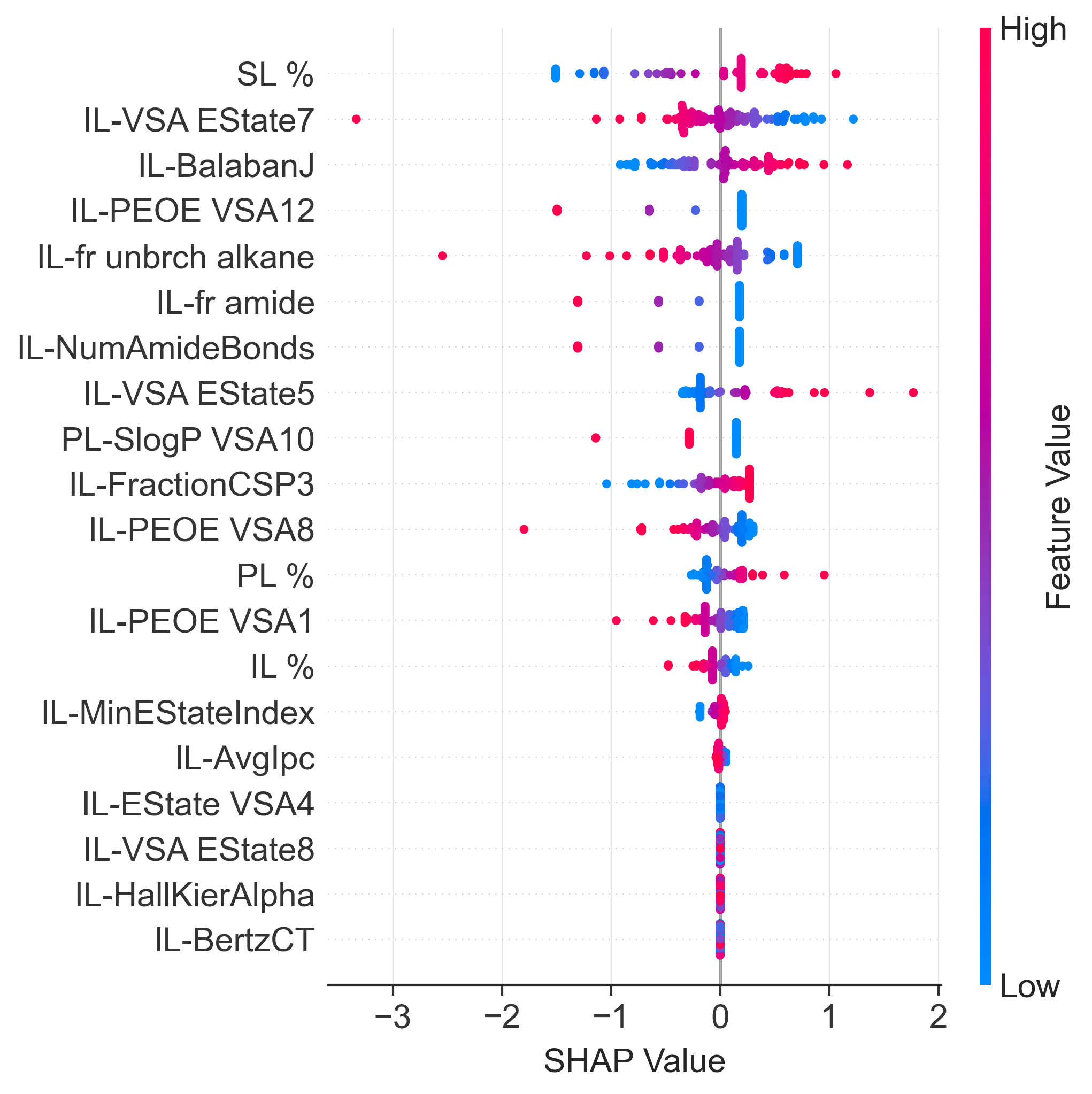}
            \caption{}
        \end{subfigure}
    \hfill
	\begin{subfigure}[t]{0.33\textwidth}
            \centering
            \includegraphics[width=0.99\textwidth]{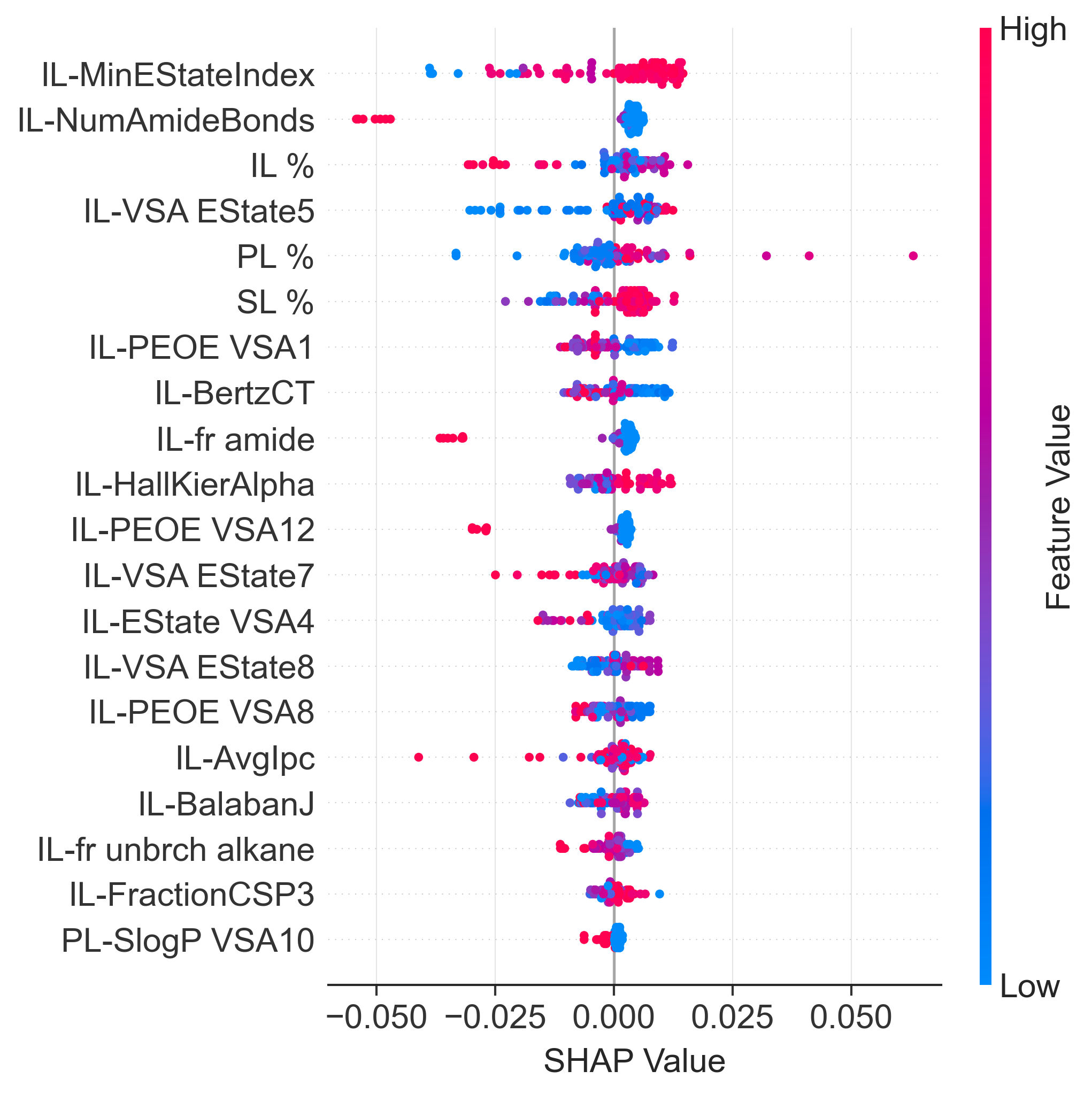}
            \caption{}
        \end{subfigure}
    \hfill
	\begin{subfigure}[t]{0.33\textwidth}
            \centering
            \includegraphics[width=0.99\textwidth]{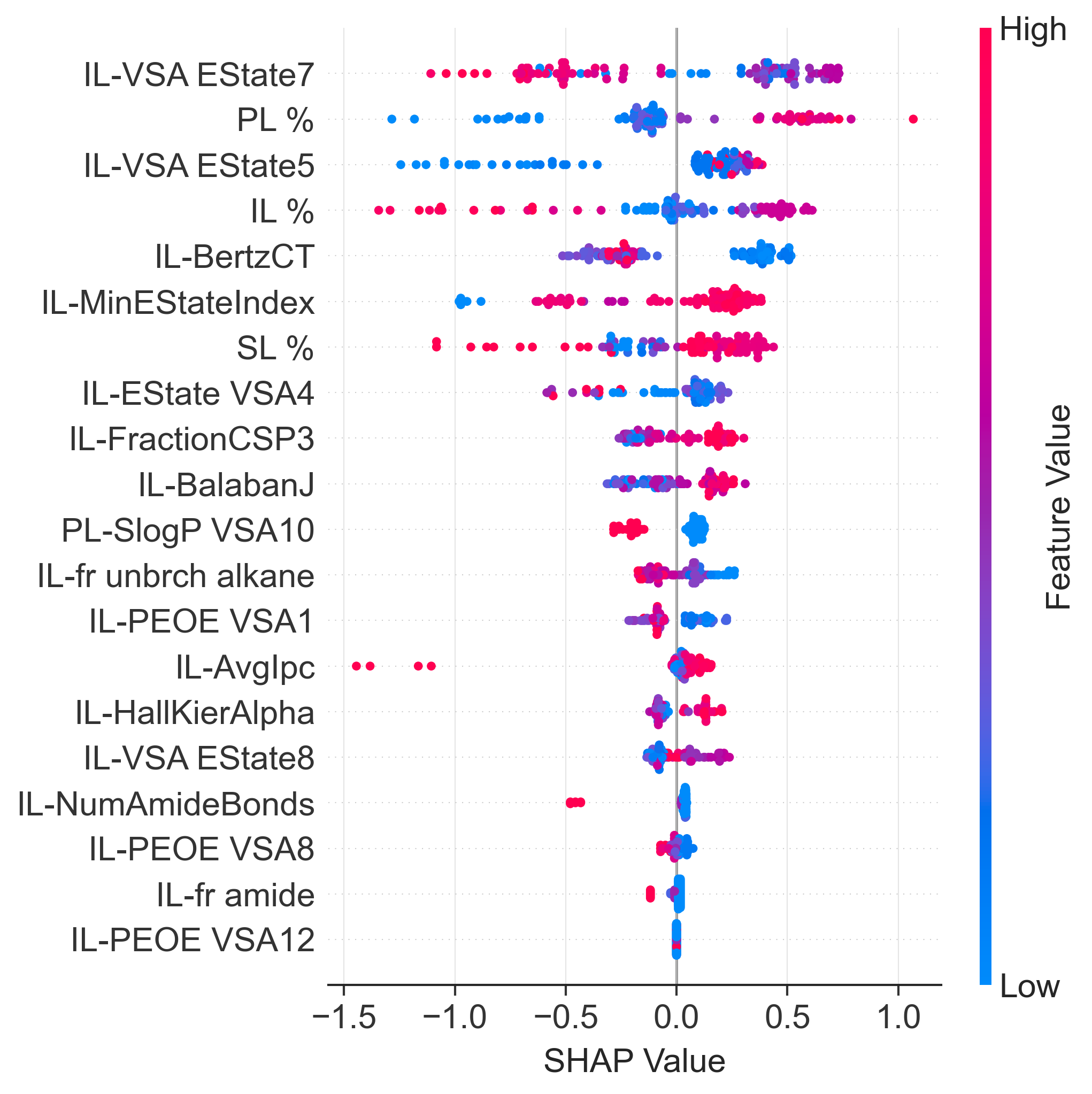}
            \caption{}
        \end{subfigure}
    \caption{\textbf{SHAP summary plots illustrating feature effects in ML models predicting LNP biodistribution.} 
    SHAP values for the top 20 consensus features are shown for (a) logistic regression, (b) random forest, and (c) XGBoost models. Each point represents a single LNP formulation in the dataset. The horizontal position of a point indicates the SHAP value, which quantifies the contribution of that feature to the model prediction for that sample; positive values increase and negative values decrease the predicted probability of liver targeting. Feature values are color-coded from low (blue) to high (pink). Feature names are prefixed by the lipid component from which the descriptor is derived (e.g., IL--), where the suffix denotes the corresponding molecular descriptor or formulation variable. The spread of points along the horizontal axis reflects how variations in each feature influence model predictions across samples, highlighting features that consistently contribute to predictions of liver-targeting behavior.}

	\label{fig:beeswarm}
\end{figure}

To further investigate how the consensus molecular determinants influenced model predictions, SHAP summary plots were generated for the top 20 consensus-ranked features identified across the three machine learning models (Figure~\ref{fig:beeswarm}). Whereas the previous analyses established which molecular descriptors were consistently important, the SHAP summary plots provide additional insight into how variations in these descriptors contributed to the classification of individual LNP formulations as liver-targeting or extrahepatic-targeting.

Despite differences in the underlying learning algorithms, the three models exhibited broadly consistent patterns in the contributions of the highest-ranked molecular determinants. In particular, random forest and XGBoost displayed remarkably similar SHAP distributions, while logistic regression identified many of the same influential descriptors despite its linear decision boundary. This agreement further supports the robustness of the consensus feature ranking and indicates that the identified molecular determinants contribute consistently to model predictions across different machine learning algorithms. 

Several of the highest-ranked descriptors, including IL-VSA EState7, IL-MinEStateIndex, and the formulation composition variables IL \%, SL \%, and PL \%, exhibited broad SHAP value distributions, indicating that variations in these features substantially influenced the classification of LNP formulations as liver-targeting or non-liver-targeting. For many of these descriptors, higher and lower feature values produced systematically different SHAP contributions, demonstrating that changes in their values were consistently associated with changes in model predictions. The broader SHAP distributions observed for these descriptors further emphasize their dominant influence on prediction relative to lower-ranked features, which generally exhibited SHAP values concentrated near zero.

The top-ranked features can be grouped into six major categories: formulation composition, electrotopological state and surface-area descriptors, charge-weighted surface-area descriptors, hydrophobicity-weighted surface-area descriptors, molecular topology and complexity descriptors, and functional group/structural motif descriptors. This categorization provides a chemically interpretable framework for understanding how formulation-level variables and molecular-level features collectively contribute to hepatic versus extrahepatic LNP accumulation and protein expression.

The mole percentages of ionizable lipid (IL\%), PEGylated/polymer-conjugated lipid (PL\%), and sterol (SL\%) were among the top-ranked features for determining biodistribution across all three models. IL\% showed the clearest extrahepatic trend in this family, with higher values associated with extrahepatic accumulation and protein expression across logistic regression, random forest, and XGBoost models. In contrast, PL\% showed the opposite trend, with higher values consistently associated with hepatic accumulation across all three models. SL\% showed weaker and more model-dependent behavior, although higher values were more hepatic in logistic regression and random forest, while XGBoost showed limited separation. These findings suggest that a higher ionizable lipid fraction, together with lower PEGylated/polymer-conjugated lipid and sterol fractions, represents a compositionally relevant signature associated with extrahepatic accumulation. The study by Zhu et al. \cite{zhu2022multi} provides independent experimental support for the importance of lipid molar composition in determining functional LNP biodistribution. Because this study was not included in our training dataset, it did not influence model learning and can therefore serve as external validation for the compositional trends identified by our ML analysis. In their multi-step high-throughput screening study, 1080 LNP formulations were initially evaluated \textit{in vitro}, after which 32 top-performing formulations for each lipid category were clustered and tested \textit{in vivo}. Among formulations composed of the same components, Dlin-MC3/DOPE/cholesterol/DMG-PEG, clusters with average molar ratios of 35/35/29.74/0.26 and 32.84/24.66/41.94/0.56 showed distinct organ-level protein expression profiles. The formulation cluster with higher ionizable lipid content and lower cholesterol and PEG-lipid fractions showed reduced hepatic protein expression (65.9\% to 49.8\%) and increased splenic expression (30.0\% to 45.2\%), consistent with our model-derived association between higher IL\%, lower PL\%, lower cholesterol, and enhanced extrahepatic delivery. This trend was further supported by individual formulations containing the same component set, Dlin-MC3/DSPC/cholesterol/DMG-PEG, but different molar ratios. Increasing the ionizable lipid fraction while decreasing cholesterol and PEG-lipid content from 36.3/3.63/59.9/0.12 to 54.5/5.45/39.9/0.08 decreased hepatic protein expression from 60.8\% to 47.9\% and increased lung expression from 2.3\% to 21.4\%. These results demonstrate that, even when lipid identities are held constant, changes in formulation composition can substantially alter functional organ-level delivery, supporting the compositional trends identified by our ML analysis.

The electrotopological surface descriptor family included IL-VSA EState5, IL-VSA EState7, IL-VSA EState8, IL-EState VSA4, and IL-MinEStateIndex. These descriptors integrate atom-level electrotopological state information with molecular surface-area contributions, capturing how distinct electronic environments are distributed across the surface of the ionizable lipid. Within this family, IL-VSA EState7 showed the strongest and most consistent class separation, with higher values associated with extrahepatic accumulation. In contrast, lower IL-VSA EState7 values were associated with hepatic accumulation in logistic regression. IL-VSA EState5 and IL-MinEStateIndex also exhibited meaningful but more model-dependent trends, with higher values more frequently associated with hepatic accumulation, whereas IL-VSA EState8 and IL-EState VSA4 showed weaker or less consistent separation between classes. Collectively, these findings suggest that electrotopological surface environments do not contribute equally to LNP biodistribution. Rather, the combination of higher IL-VSA EState7 with lower IL-VSA EState5 and IL-MinEStateIndex may define an ionizable-lipid surface profile associated with improved extrahepatic accumulation and functional protein expression. Chemically, this descriptor pattern may reflect a heteroatom-rich headgroup or linker region in which tertiary amine, ester, amide, or ether functionalities create a distinct surface-exposed electronic environment. For example, atoms adjacent to an ionizable tertiary amine and nearby carbonyl or ether oxygens may exhibit altered electrotopological states due to differences in electronegativity, bonding pattern, and molecular connectivity. Thus, high IL-VSA EState7 likely reflects a specific arrangement of heteroatom-proximal surface atoms rather than the presence of a single functional group, while lower IL-VSA EState5 and IL-MinEStateIndex suggest limited contribution from alternative lower-EState surface environments or localized electronically perturbed regions within the ionizable lipid structure.

Charge-weighted surface-area descriptors included IL-PEOE VSA1, IL-PEOE VSA8, and IL-PEOE VSA12. These descriptors group atoms according to partial charge ranges and sum their approximate surface-area contributions, thereby capturing how charged or weakly charged atomic environments are distributed across the ionizable lipid surface. Within this family, IL-PEOE VSA1 showed the most informative and consistent pattern. Higher IL-PEOE VSA1 values were associated with extrahepatic accumulation in logistic regression, whereas lower values were associated with hepatic accumulation in random forest and XGBoost. In contrast, IL-PEOE VSA8 and IL-PEOE VSA12 showed weaker and more model-dependent separation between hepatic and extrahepatic classes. These findings suggest that the distribution of partial charge across the ionizable lipid surface may contribute to LNP biodistribution. In particular, higher IL-PEOE VSA1 may reflect greater surface contribution from strongly electronegative atomic environments, such as oxygen- or nitrogen-containing regions associated with ester, ether, amide, carbonyl, or tertiary amine-containing motifs. Chemically, this descriptor may capture ionizable lipids in which heteroatom-rich headgroup or linker regions create localized charged or polar surface domains that influence interactions with serum proteins, cell membranes, or endosomal interfaces. The weaker trends observed for IL-PEOE VSA8 and IL-PEOE VSA12 indicate that not all charge-defined surface regions are equally informative. Rather, specific partial-charge bins, particularly those captured by IL-PEOE VSA1, may be more closely associated with extrahepatic accumulation and functional protein expression.

The hydrophobicity-weighted surface-area category was represented by PL-SlogP VSA10, a descriptor derived from the PEGylated/polymer-conjugated lipid. SlogP VSA descriptors group atoms according to atomic logP contributions and sum their corresponding surface-area contributions. PL-SlogP VSA10 was one of the clearest non-ionizable lipid features associated with extrahepatic accumulation: higher values were associated with extrahepatic accumulation across logistic regression, random forest, and XGBoost, while lower values were more hepatic in XGBoost. This suggests that the hydrophobic surface contribution of the PEGylated/polymer-conjugated lipid, likely related to the lipid anchor or polymer–lipid interface, may be an important determinant of LNP biodistribution. This is consistent with the known role of PEG/polymer-lipid hydrophobic anchoring in controlling surface presentation and PEG/polymer retention within the nanoparticle \cite{silvius1993interbilayer, semple2005immunogenicity}. Mui et al. \cite{mui2013influence} demonstrated that approximately 80\% of the initially incorporated \textsuperscript{3}H-labeled PEG-C14 lipid dissociated from LNPs within 2 h in mouse plasma, whereas PEG-C16 and PEG-C18 lipids were retained for substantially longer periods. This difference was attributed to the greater energetic barrier required for longer hydrophobic chains to desorb from the lipid membrane. Consistent with this mechanism, the estimated desorption rates were 45\%, 1.3\%, and 0.2\% per hour for PEG-C14, PEG-C16, and PEG-C18 lipids, respectively. The authors further showed that PEG-lipid anchor length influenced hepatic accumulation. PEG-C14-containing LNPs accumulated rapidly in the liver, reaching approximately 55\% of the injected dose at 4 h. In contrast, PEG-C16- and PEG-C18-containing LNPs accumulated more slowly and reached lower maximum hepatic levels of approximately 35\% and 25\% of the injected dose, respectively.

Molecular topology and complexity descriptors included IL-BertzCT, IL-BalabanJ, IL-HallKierAlpha, IL-FractionCSP3, and IL-AvgIpc. These descriptors capture global structural characteristics of the ionizable lipid, including molecular complexity, graph compactness, molecular shape, degree of saturation, and graph-based information content. Within this family, IL-FractionCSP3 showed the clearest and most consistent hepatic trend, with higher values associated with hepatic accumulation across all three models and lower values associated with extrahepatic accumulation in logistic regression. Because FractionCSP3 reflects the proportion of sp\textsuperscript{3}-hybridized carbons, higher values generally indicate a more saturated, aliphatic, and three-dimensional carbon framework, whereas lower values suggest greater contribution from unsaturated, sp\textsuperscript{2}-rich, or more conformationally restricted structural elements. IL-BalabanJ and IL-HallKierAlpha also showed hepatic trends at higher values, particularly in logistic regression/XGBoost and random forest/XGBoost, respectively. These descriptors reflect aspects of molecular connectivity, compactness, branching, atom-type composition, and shape. Thus, higher values may capture ionizable lipids with more compact or topology-dependent structural arrangements that favor hepatic accumulation. In contrast, IL-BertzCT showed an opposite pattern in the tree-based models, where lower values were associated with hepatic accumulation, suggesting that greater molecular complexity may be more favorable for extrahepatic accumulation in nonlinear models. IL-AvgIpc showed limited class separation, indicating that not all graph-information descriptors contribute equally to biodistribution prediction. Overall, the strongest conclusion from this descriptor family is that ionizable lipids with high IL-FractionCSP3, and to a lesser extent high IL-BalabanJ and IL-HallKierAlpha, are more consistently associated with hepatic accumulation than extrahepatic delivery. Chemically, this suggests that highly saturated, aliphatic, and topology-defined ionizable lipid frameworks may favor liver accumulation, whereas lower saturation and/or greater structural complexity may contribute to extrahepatic accumulation and functional protein expression. However, because these descriptors capture global molecular topology rather than discrete functional groups, their interpretation should be viewed as structural guidance rather than a direct mechanistic assignment.

Functional group and structural motif descriptors included IL-NumAmideBonds, IL-fr amide, and IL-fr unbrch alkane. IL-NumAmideBonds and IL-fr amide quantify amide-containing motifs within the ionizable lipid, whereas IL-fr unbrch alkane captures the presence of linear hydrophobic alkyl segments. Within this family, the amide-related descriptors showed a consistent pattern in which low amide content was associated with hepatic accumulation, while higher IL-fr amide values were associated with extrahepatic accumulation in random forest and XGBoost. This suggests that amide-containing structural motifs may contribute favorably to extrahepatic delivery, potentially by altering hydrogen-bonding capacity, polarity, lipid packing, degradability, or interactions with proteins and biological membranes. IL-fr unbrch alkane also showed an extrahepatic trend, with higher values associated with extrahepatic accumulation in logistic regression and random forest, and lower values associated with hepatic accumulation in XGBoost. Chemically, this descriptor may reflect defined linear hydrophobic domains within the ionizable lipid tails or linker regions. Such unbranched alkyl segments can influence lipid packing, membrane insertion, nanoparticle stability, and interactions with biological interfaces. Together, these findings suggest that ionizable lipids containing both amide motifs and well-defined linear hydrophobic alkyl segments may be more likely to support extrahepatic accumulation and functional protein expression. However, because these descriptors capture structural motifs rather than complete mechanistic pathways, their interpretation should be viewed as design guidance for prioritizing ionizable lipid chemistries rather than as direct evidence of causality.

Taken together, the most differentiating features associated with extrahepatic accumulation were higher IL\%, lower PL\%, higher IL-VSA EState7, higher PL-SlogP VSA10, higher IL-fr amide, and higher IL-fr unbrch alkane. In contrast, higher PL\%, higher IL-FractionCSP3, higher IL-BalabanJ, higher IL-HallKierAlpha, and higher IL-MinEStateIndex were more frequently associated with hepatic accumulation. Based on these trends, an LNP formulation designed for extrahepatic accumulation could prioritize a relatively high ionizable lipid fraction, avoid excessive PEGylated/polymer-conjugated lipid content, incorporate PEG/polymer-lipid structures with favorable hydrophobic anchor-associated surface features, and use ionizable lipids enriched in the molecular surface/electronic environments captured by IL-VSA EState7. At the ionizable lipid level, candidate structures could preferentially include amide-containing motifs and appropriate unbranched alkyl segments while avoiding structural profiles associated with high FractionCSP3, high BalabanJ, high HallKierAlpha, or high MinEStateIndex. These recommendations provide interpretable, descriptor-guided design principles for prioritizing LNP formulations with a higher probability of extrahepatic accumulation.

Collectively, these results demonstrate that the consensus molecular determinants not only contain strong predictive information but also contribute to model predictions in a consistent and interpretable manner across different machine learning algorithms. The close agreement between the SHAP analyses of logistic regression, random forest, and XGBoost provides additional confidence that the identified molecular determinants represent robust predictors of LNP biodistribution rather than artifacts of a specific machine learning model.


\section{Conclusion}

This work demonstrates that interpretable machine learning can bridge a critical knowledge gap in LNP design by linking molecular structure and formulation composition to in vivo biodistribution. By curating a diverse literature-derived dataset of 476 LNP formulations and constructing an 808-dimensional molecular and compositional feature representation, we show that hepatic versus extrahepatic accumulation can be predicted with strong accuracy across distinct learning paradigms. The superior performance of random forest and XGBoost relative to logistic regression highlights the nonlinear nature of LNP biodistribution, while the competitive performance of all three models confirms that chemically meaningful descriptors contain substantial predictive information.

Beyond prediction, the central contribution of this study is interpretability. SHAP analysis and consensus feature ranking identified a relatively small set of molecular determinants that accounted for much of the model's predictive behavior. These features reveal that extrahepatic accumulation is not governed by ionizable lipid chemistry alone, but by the combined influence of ionizable-lipid surface electronics, charge distribution, molecular topology, amide and alkyl motifs, PEG/polymer-lipid hydrophobic surface features, and formulation composition. In particular, higher ionizable lipid fraction, lower PEGylated/polymer-conjugated lipid fraction, favorable IL-VSA EState7 values, PL-SlogP VSA10, and ionizable-lipid amide and unbranched alkane motifs emerged as practical design-relevant features.

These findings provide a rational framework for prioritizing LNP formulations with increased probability of extrahepatic accumulation before resource-intensive in vivo screening. More broadly, this study illustrates how interpretable machine learning can convert heterogeneous literature data into actionable nanomedicine design rules. Expanding this framework with larger organ-resolved datasets, standardized biodistribution measurements, and broader chemical diversity across helper, sterol, PEGylated, and polymer-conjugated lipids will further improve generalizability and accelerate the design of tissue-selective RNA delivery systems.

\small
\bibliographystyle{unsrt}
\bibliography{references.bib}

\clearpage
\renewcommand{\thefigure}{S\arabic{figure}}  
\setcounter{figure}{0}

\renewcommand{\thetable}{S\arabic{table}}  
\setcounter{table}{0}

\section*{Supplementary Information}

\begin{table*}[ht]
\centering
\caption{\textbf{Literature sources used to construct the curated LNP dataset.}
The table lists the publications from which LNP formulations were extracted during dataset curation. The numbers indicate how many LNPs were extracted from each study. The final row reports the total number of LNPs included in the dataset.}
\label{tab:dataset_sources}

\begin{tabular}{lc@{\hspace{1.7cm}}lc}
\toprule
\textbf{DOI} & \textbf{LNPs} & \textbf{DOI} & \textbf{LNPs} \\
\midrule

\href{https://doi.org/10.1039/d1bm00866h}{10.1039/D1BM00866H} & 4 &
\href{https://doi.org/10.1016/j.jconrel.2019.10.028}{10.1016/j.jconrel.2019.10.028} & 6 \\

\href{https://doi.org/10.1039/d2bm00168c}{10.1039/D2BM00168C} & 3 &
\href{https://doi.org/10.1038/s42003-021-02441-2}{10.1038/s42003-021-02441-2} & 4 \\

\href{https://doi.org/10.1016/j.jconrel.2016.05.059}{10.1016/j.jconrel.2016.05.059} & 3 &
\href{https://doi.org/10.1021/acs.nanolett.5b02497}{10.1021/acs.nanolett.5b02497} & 1 \\

\href{https://doi.org/10.1016/j.ijpharm.2016.06.124}{10.1016/j.ijpharm.2016.06.124} & 1 &
\href{https://doi.org/10.1021/acs.nanolett.0c00596}{10.1021/acs.nanolett.0c00596} & 5 \\

\href{https://doi.org/10.1002/smll.201805097}{10.1002/smll.201805097} & 11 &
\href{https://doi.org/10.1016/j.jconrel.2013.09.027}{10.1016/j.jconrel.2013.09.027} & 5 \\

\href{https://doi.org/10.1126/sciadv.abf4398}{10.1126/sciadv.abf4398} & 5 &
\href{https://doi.org/10.1016/j.biomaterials.2012.05.002}{10.1016/j.biomaterials.2012.05.002} & 1 \\

\href{https://doi.org/10.1039/c9nr05788a}{10.1039/C9NR05788A} & 11 &
\href{https://doi.org/10.1002/smll.202304378}{10.1002/smll.202304378} & 2 \\

\href{https://doi.org/10.1002/anie.201809055}{10.1002/anie.201809055} & 2 &
\href{https://doi.org/10.1021/acsanm.0c01834}{10.1021/acsanm.0c01834} & 6 \\

\href{https://doi.org/10.1039/d3tb00516j}{10.1039/D3TB00516J} & 3 &
\href{https://doi.org/10.1016/j.ijpharm.2022.122489}{10.1016/j.ijpharm.2022.122489} & 1 \\

\href{https://doi.org/10.1002/ange.202013927}{10.1002/ange.202013927} & 10 &
\href{https://doi.org/10.1126/sciadv.aba1028}{10.1126/sciadv.aba1028} & 14 \\

\href{https://doi.org/10.1038/s41565-020-0669-6}{10.1038/s41565-020-0669-6} & 8 &
\href{https://doi.org/10.1016/j.ymthe.2021.06.004}{10.1016/j.ymthe.2021.06.004} & 1 \\

\href{https://doi.org/10.1073/pnas.2307813120}{10.1073/pnas.2307813120} & 6 &
\href{https://doi.org/10.1021/jacs.2c12893}{10.1021/jacs.2c12893} & 3 \\

\href{https://doi.org/10.1073/pnas.2311276120}{10.1073/pnas.2311276120} & 1 &
\href{https://doi.org/10.1126/sciadv.ade1444}{10.1126/sciadv.ade1444} & 2 \\

\href{https://doi.org/10.1002/anie.202310401}{10.1002/anie.202310401} & 65 &
\href{https://doi.org/10.1016/j.jconrel.2024.05.015}{10.1016/j.jconrel.2024.05.015} & 5 \\

\href{https://doi.org/10.1016/j.jconrel.2021.01.005}{10.1016/j.jconrel.2021.01.005} & 3 &
\href{https://doi.org/10.1002/adfm.202303795}{10.1002/adfm.202303795} & 29 \\

\href{https://doi.org/10.1038/s41467-020-17029-3}{10.1038/s41467-020-17029-3} & 4 &
\href{https://doi.org/10.1021/acs.nanolett.3c03509}{10.1021/acs.nanolett.3c03509} & 5 \\

\href{https://doi.org/10.1021/acs.nanolett.3c05031}{10.1021/acs.nanolett.3c05031} & 2 &
\href{https://doi.org/10.1073/pnas.2307809121}{10.1073/pnas.2307809121} & 1 \\

\href{https://doi.org/10.1038/s41467-022-35637-z}{10.1038/s41467-022-35637-z} & 2 &
\href{https://doi.org/10.1038/s41598-024-57997-w}{10.1038/s41598-024-57997-w} & 1 \\

\href{https://doi.org/10.1016/j.colsurfb.2024.113980}{10.1016/j.colsurfb.2024.113980} & 2 &
\href{https://doi.org/10.1002/adhm.201901487}{10.1002/adhm.201901487} & 21 \\

\href{https://doi.org/10.1016/j.jconrel.2021.11.022}{10.1016/j.jconrel.2021.11.022} & 4 &
\href{https://doi.org/10.1073/pnas.2020401118}{10.1073/pnas.2020401118} & 3 \\

\href{https://doi.org/10.1002/adfm.202312038}{10.1002/adfm.202312038} & 6 &
\href{https://doi.org/10.1016/j.nantod.2024.102325}{10.1016/j.nantod.2024.102325} & 2 \\

\href{https://doi.org/10.1002/adma.201606944}{10.1002/adma.201606944} & 1 &
\href{https://doi.org/10.34133/bmr.0017}{10.34133/bmr.0017} & 2 \\

\href{https://doi.org/10.1016/j.jconrel.2020.06.030}{10.1016/j.jconrel.2020.06.030} & 10 &
\href{https://doi.org/10.1016/j.jconrel.2024.04.018}{10.1016/j.jconrel.2024.04.018} & 3 \\

\href{https://doi.org/10.1016/j.jconrel.2022.03.046}{10.1016/j.jconrel.2022.03.046} & 18 &
\href{https://doi.org/10.1016/j.omtn.2020.01.018}{10.1016/j.omtn.2020.01.018} & 3 \\

\href{https://doi.org/10.1021/acs.nanolett.2c03234}{10.1021/acs.nanolett.2c03234} & 3 &
\href{https://doi.org/10.1002/anie.202302676}{10.1002/anie.202302676} & 1 \\

\href{https://doi.org/10.1038/s41563-020-00886-0}{10.1038/s41563-020-00886-0} & 31 &
\href{https://doi.org/10.1016/j.jconrel.2024.05.036}{10.1016/j.jconrel.2024.05.036} & 1 \\

\href{https://doi.org/10.1021/acsnano.4c01171}{10.1021/acsnano.4c01171} & 3 &
\href{https://doi.org/10.1021/jacs.3c09143}{10.1021/jacs.3c09143} & 16 \\

\href{https://doi.org/10.1002/adma.202302901}{10.1002/adma.202302901} & 12 &
\href{https://doi.org/10.1038/s41467-024-45422-9}{10.1038/s41467-024-45422-9} & 4 \\

\href{https://doi.org/10.1002/smll.202303568}{10.1002/smll.202303568} & 3 &
\href{https://doi.org/10.1039/d1tb00736j}{10.1039/D1TB00736J} & 1 \\

\href{https://doi.org/10.1021/jacs.3c05574}{10.1021/jacs.3c05574} & 4 &
\href{https://doi.org/10.1021/jacs.4c04565}{10.1021/jacs.4c04565} & 4 \\

\href{https://doi.org/10.1021/acs.nanolett.4c01235}{10.1021/acs.nanolett.4c01235} & 5 &
\href{https://doi.org/10.1002/adma.202303614}{10.1002/adma.202303614} & 1 \\

\href{https://doi.org/10.1039/d1bm01454d}{10.1039/D1BM01454D} & 14 &
\href{https://doi.org/10.1002/anie.202013927}{10.1002/anie.202013927} & 20 \\

\href{https://doi.org/10.1002/adma.201805308}{10.1002/adma.201805308} & 4 &
\href{https://doi.org/10.1002/jbm.a.37705}{10.1002/jbm.a.37705} & 5 \\

\href{https://doi.org/10.1038/s41565-023-01404-4}{10.1038/s41565-023-01404-4} & 1 &
\href{https://doi.org/10.1002/advs.202202556}{10.1002/advs.202202556} & 1 \\

\href{https://doi.org/10.1016/j.ymthe.2019.03.001}{10.1016/j.ymthe.2019.03.001} & 1 &
\href{https://doi.org/10.1021/acsnano.3c06225}{10.1021/acsnano.3c06225} & 1 \\

\href{https://doi.org/10.1039/d2bm01846b}{10.1039/D2BM01846B} & 3 &
\href{https://doi.org/10.1038/mt.2011.141}{10.1038/mt.2011.141} & 3 \\

\href{https://doi.org/10.1093/nsr/nwae135}{10.1093/nsr/nwae135} & 3 &
\href{https://doi.org/10.1002/anie.202008082}{10.1002/anie.202008082} & 4 \\

\href{https://doi.org/10.1101/2022.12.22.521490}{10.1101/2022.12.22.521490} & 3 &
\href{https://doi.org/10.1021/acscentsci.4c00071}{10.1021/acscentsci.4c00071} & 1 \\

\href{https://doi.org/10.3390/ijms25031388}{10.3390/ijms25031388} & 3 &
\href{https://doi.org/10.1007/s12274-018-2082-0}{10.1007/s12274-018-2082-0} & 1 \\

\href{https://doi.org/10.1021/acsnano.1c04456}{10.1021/acsnano.1c04456} & 1 &
\href{https://doi.org/10.1002/smll.202105832}{10.1002/smll.202105832} & 1 \\

\href{https://doi.org/10.1038/s41557-024-01557-2}{10.1038/s41557-024-01557-2} & 6 & \\

\midrule
\multicolumn{3}{r}{\textbf{Total}} & \textbf{476} \\
\bottomrule
\end{tabular}

\end{table*}

\begin{figure}[ht]
	\centering
	\begin{subfigure}[t]{0.33\textwidth}
            \centering
            \includegraphics[width=0.99\textwidth]{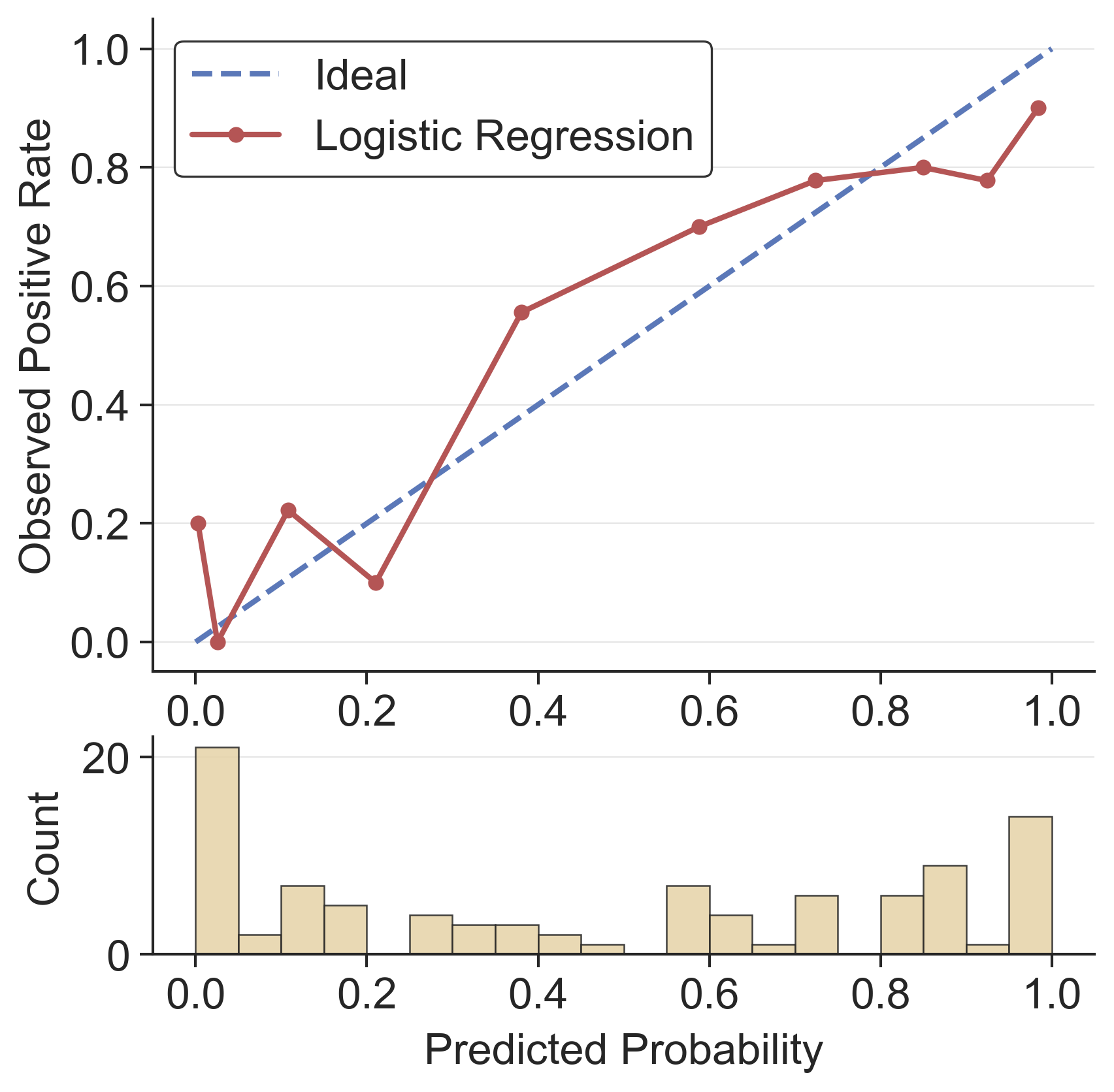}
            \caption{}
        \end{subfigure}
    \hfill
	\begin{subfigure}[t]{0.33\textwidth}
            \centering
            \includegraphics[width=0.99\textwidth]{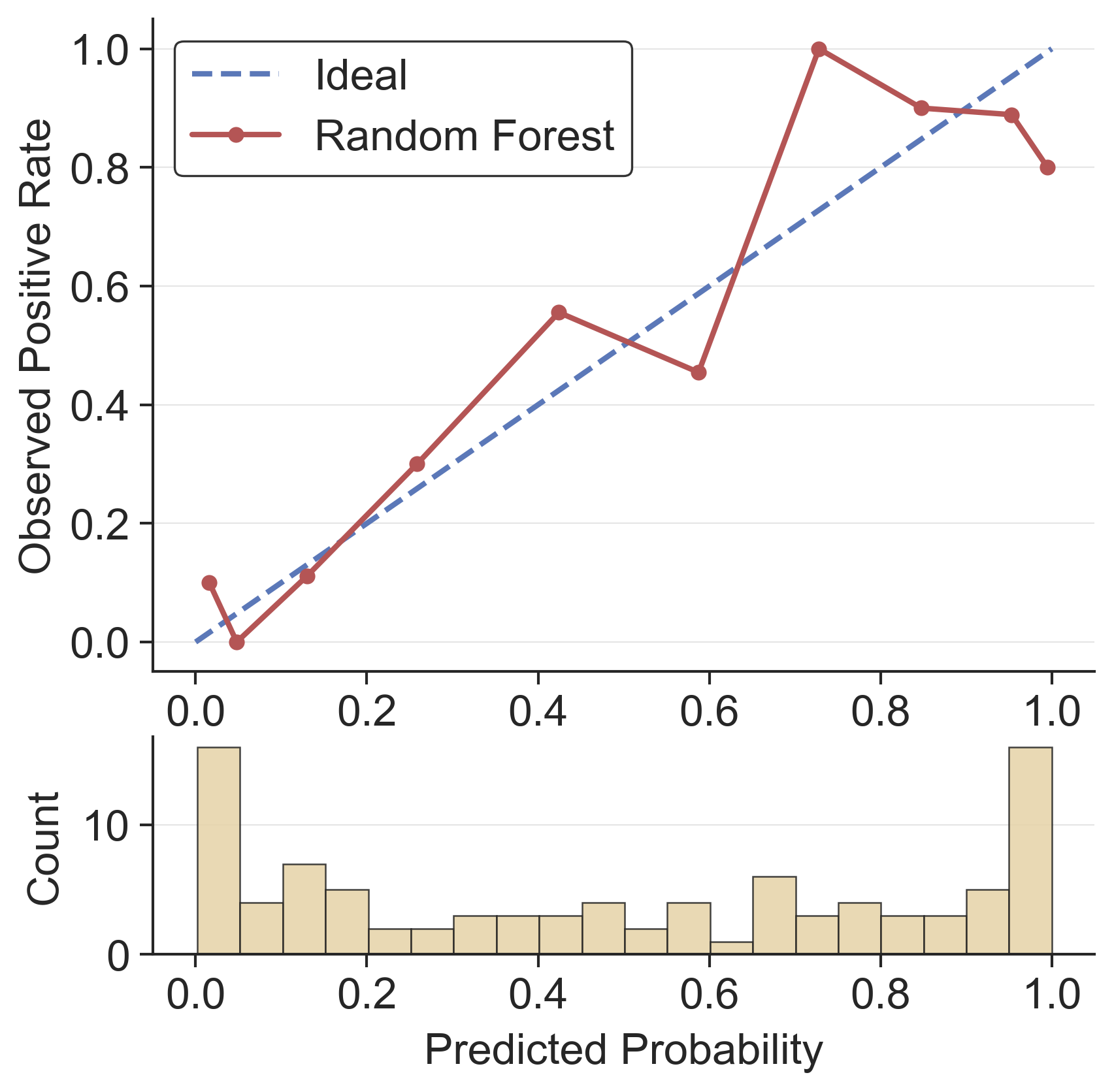}
            \caption{}
        \end{subfigure}
    \hfill
	\begin{subfigure}[t]{0.33\textwidth}
            \centering
            \includegraphics[width=0.99\textwidth]{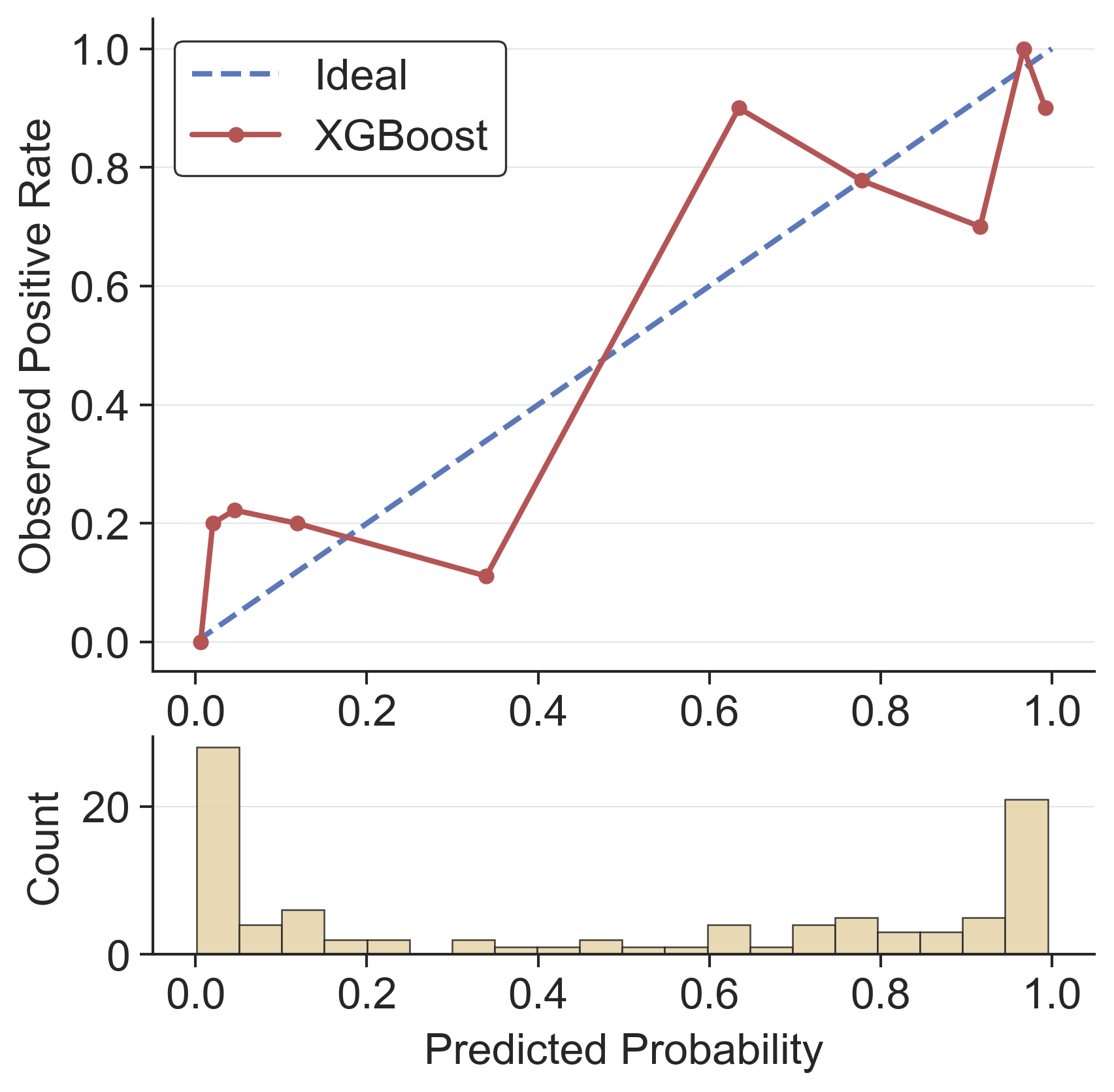}
            \caption{}
        \end{subfigure}
    \caption{\textbf{Calibration analysis of ML models predicting LNP biodistribution.}
    Calibration curves for (a) logistic regression, (b) random forest, and (c) XGBoost models are shown. The upper panels compare observed positive rates with predicted probabilities across probability bins; the dashed diagonal line indicates ideal calibration where predicted probabilities perfectly match observed frequencies. The lower panels show histograms of predicted probabilities for the corresponding models, illustrating the distribution of model confidence across the test samples.}

	\label{fig:calib}
\end{figure}

\begin{table}[ht]
\caption{\textbf{Calibration performance of the evaluated machine learning models trained using the full feature set.} Lower Brier scores indicate better calibration, corresponding to better agreement between predicted probabilities and observed outcomes.}
\label{tab:supp_brier_scores}
\centering
\begin{tabular}{lc}
\toprule
\textbf{Model} & \textbf{Brier Score} \\
\midrule
logistic regression & 0.160 \\
random forest       & 0.147 \\
XGBoost             & \textbf{0.144} \\
\bottomrule
\end{tabular}
\end{table}

\begin{table}[ht]
\centering
\caption{\textbf{Top 50 consensus-ranked features identified across ML models.} 
Feature rankings from logistic regression (LR), random forest (RF), and XGBoost (XGB) are shown along with the median rank used to aggregate model-specific rankings. The consensus rank represents the final ordering obtained after sorting features by median rank and resolving ties using the average normalized SHAP importance values.}
\label{tab:top50_consensus_features}

\begin{tabular}{c l c c c c}
\toprule
\textbf{Consensus Rank} & \textbf{Feature} & \textbf{LR Rank} & \textbf{RF Rank} & \textbf{XGB Rank} & \textbf{Median Rank} \\
\midrule
1  & IL-VSA EState5        & 17   & 5   & 4   & 5 \\
2  & IL \%                 & 59   & 4   & 5   & 5 \\
3  & IL-VSA EState7        & 6    & 20  & 1   & 6 \\
4  & PL \%                 & 36   & 6   & 3   & 6 \\
5  & SL \%                 & 1    & 7   & 9   & 7 \\
6  & IL-MinEStateIndex     & 148  & 1   & 7   & 7 \\
7  & IL-BertzCT            & 735  & 9   & 6   & 9 \\
8  & IL-BalabanJ           & 7    & 35  & 15  & 15 \\
9  & IL-NumAmideBonds      & 15.5 & 2   & 36  & 15.5 \\
10 & IL-fr amide           & 15.5 & 11  & 92  & 15.5 \\
11 & IL-PEOE VSA12         & 8    & 18  & 525.5 & 18 \\
12 & IL-fr unbrch alkane   & 12   & 69  & 22  & 22 \\
13 & IL-PEOE VSA1          & 38   & 8   & 23  & 23 \\
14 & IL-EState VSA4        & 735  & 24  & 12  & 24 \\
15 & PL-SlogP VSA10        & 26   & 117 & 17  & 26 \\
16 & IL-HallKierAlpha      & 735  & 14  & 26  & 26 \\
17 & IL-FractionCSP3       & 27   & 84  & 13  & 27 \\
18 & IL-AvgIpc             & 161.5 & 27 & 24  & 27 \\
19 & IL-VSA EState8        & 735  & 25  & 27  & 27 \\
20 & IL-PEOE VSA8          & 28   & 26  & 87  & 28 \\
21 & IL-SPS                & 30   & 36  & 14  & 30 \\
22 & IL-EState VSA2        & 80   & 30  & 21  & 30 \\
23 & IL-BCUT2D MRHI        & 5    & 31  & 74  & 31 \\
24 & IL-SMR VSA3           & 89   & 3   & 31  & 31 \\
25 & IL-BCUT2D MWHI        & 2    & 33  & 69  & 33 \\
26 & IL-EState VSA5        & 175  & 34  & 28  & 34 \\
27 & IL-SlogP VSA1         & 117  & 13  & 35  & 35 \\
28 & IL-PEOE VSA6          & 35   & 101 & 34  & 35 \\
29 & IL-SlogP VSA3         & 735  & 16  & 37  & 37 \\
30 & IL-BCUT2D MRLOW       & 64   & 39  & 19  & 39 \\
31 & IL-VSA EState2        & 67   & 12  & 41  & 41 \\
32 & IL-MaxPartialCharge   & 77.5 & 42  & 40  & 42 \\
33 & IL-BCUT2D MWLOW       & 68   & 44  & 11  & 44 \\
34 & IL-Phi                & 103  & 15  & 44  & 44 \\
35 & IL-NOCount            & 46   & 40  & 525.5 & 46 \\
36 & IL-Chi4n              & 735  & 41  & 46  & 46 \\
37 & IL-BCUT2D LOGPLOW     & 138  & 22  & 47  & 47 \\
38 & IL-BCUT2D CHGLO       & 735  & 48  & 10  & 48 \\
39 & IL-Kappa3             & 735  & 51  & 20  & 51 \\
40 & IL-SMR VSA5           & 735  & 52  & 50  & 52 \\
41 & IL-MaxAbsEStateIndex  & 164.5 & 54 & 33  & 54 \\
42 & IL-EState VSA7        & 324  & 55  & 25  & 55 \\
43 & IL-BCUT2D LOGPHI      & 735  & 19  & 56  & 56 \\
44 & IL-EState VSA3        & 735  & 56  & 45  & 56 \\
45 & IL-EState VSA1        & 735  & 10  & 57  & 57 \\
46 & IL-VSA EState9        & 84   & 58  & 18  & 58 \\
47 & IL-BCUT2D CHGHI       & 75   & 17  & 60  & 60 \\
48 & IL-FpDensityMorgan3   & 63   & 60  & 68  & 63 \\
49 & IL-SlogP VSA5         & 735  & 63  & 39  & 63 \\
50 & HL \%                 & 735  & 64  & 53  & 64 \\
\bottomrule
\end{tabular}
\end{table}

\begin{table*}[ht]
\centering
\caption{\textbf{Extended performance comparison of ML models across different feature-set sizes for LNP classification.}
This table extends Table~1 by including additional reduced feature sets containing the top 40 and top 5 features. Models were evaluated using the full feature set (\textit{All}) and reduced feature sets containing the top 40, 20, 10, and 5 features. Cross-validation performance is reported as the mean AUC $\pm$ standard deviation across folds, and held-out test performance is reported using AUC, accuracy, recall, and F1-score. The best AUC values within each feature-set block are highlighted in bold.}
\label{tab:model_performance_supp}

\begin{tabular}{llccccc}
\toprule
\textbf{Feature Set} & \textbf{Model} & \textbf{CV AUC} & \textbf{Test AUC} & \textbf{Accuracy} & \textbf{Recall} & \textbf{F1-score} \\
\midrule

\multirow{3}{*}{All}
& logistic regression & $0.826 \pm 0.054$ & 0.839 & 0.792 & 0.792 & 0.792 \\
& random forest       & $\mathbf{0.882} \pm 0.036$ & 0.866 & 0.802 & 0.792 & 0.800 \\
& XGBoost             & $0.875 \pm 0.034$ & \textbf{0.874} & 0.854 & 0.854 & 0.854 \\
\midrule

\multirow{3}{*}{40}
& logistic regression & $0.787 \pm 0.055$ & 0.775 & 0.698 & 0.667 & 0.688 \\
& random forest       & $0.846 \pm 0.047$ & \textbf{0.872} & 0.771 & 0.792 & 0.776 \\
& XGBoost             & $\mathbf{0.848} \pm 0.048$ & 0.858 & 0.781 & 0.812 & 0.788 \\
\midrule

\multirow{3}{*}{20}
& logistic regression & $0.759 \pm 0.056$ & 0.724 & 0.615 & 0.583 & 0.602 \\
& random forest       & $\mathbf{0.854} \pm 0.050$ & \textbf{0.866} & 0.781 & 0.812 & 0.788 \\
& XGBoost             & $0.852 \pm 0.048$ & 0.851 & 0.781 & 0.812 & 0.788 \\
\midrule

\multirow{3}{*}{10}
& logistic regression & $0.707 \pm 0.053$ & 0.714 & 0.729 & 0.750 & 0.735 \\
& random forest       & $0.842 \pm 0.054$ & \textbf{0.859} & 0.760 & 0.792 & 0.768 \\
& XGBoost             & $\mathbf{0.852} \pm 0.048$ & 0.847 & 0.771 & 0.771 & 0.771 \\
\midrule

\multirow{3}{*}{5}
& logistic regression & $0.670 \pm 0.046$ & 0.677 & 0.615 & 0.500 & 0.565 \\
& random forest       & $0.795 \pm 0.070$ & \textbf{0.819} & 0.740 & 0.771 & 0.747 \\
& XGBoost             & $\mathbf{0.807} \pm 0.063$ & 0.805 & 0.719 & 0.771 & 0.733 \\
\bottomrule

\end{tabular}
\end{table*}

\begin{figure}[ht]
	\centering
        \includegraphics[width=0.65\textwidth]{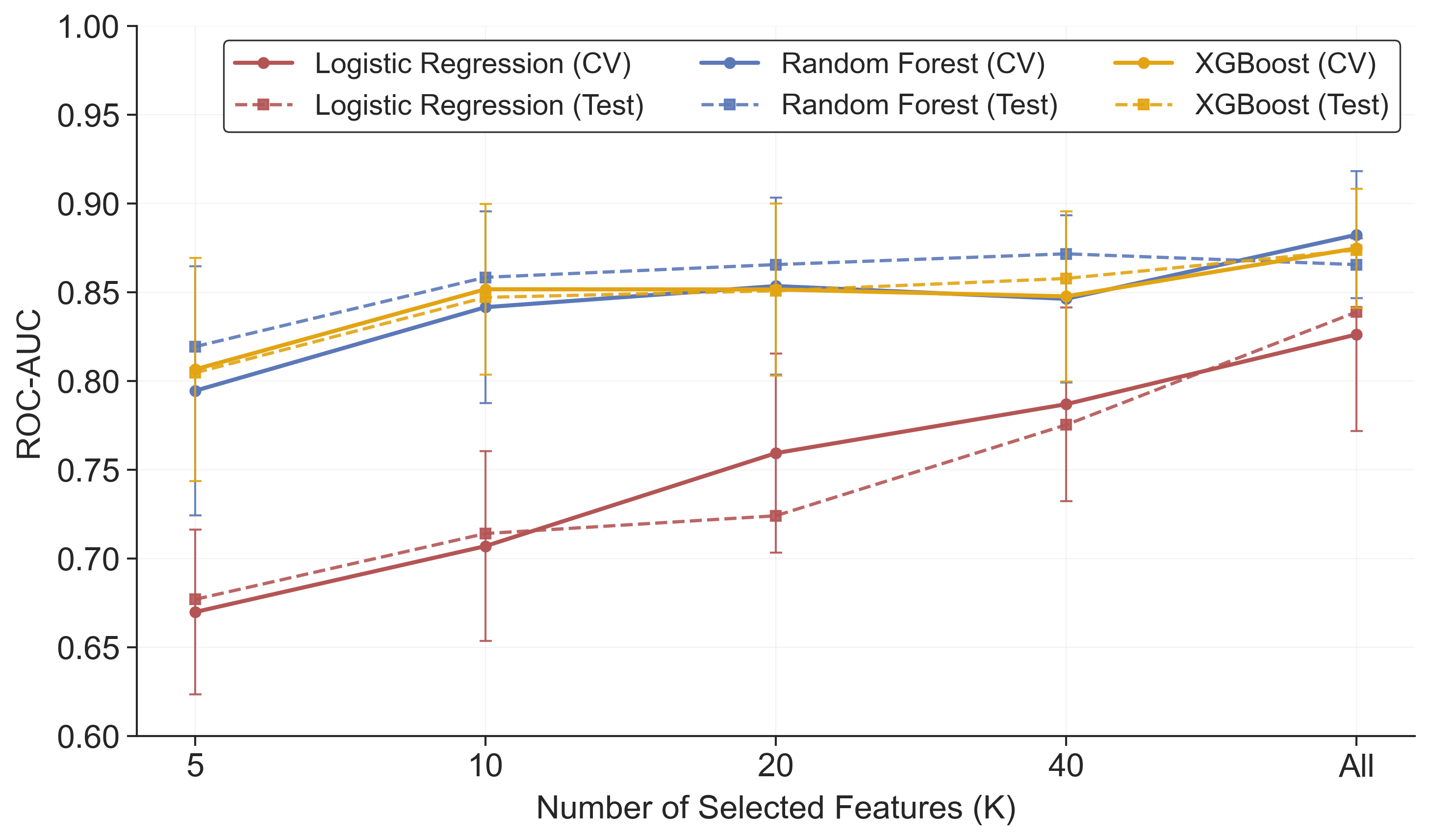}
        \caption{\textbf{Model performance as a function of the number of selected features.} 
        Model performance (ROC–AUC) is shown for logistic regression, random forest, and XGBoost models using different numbers of selected features ($K=5, 10, 20, 40$, and all features). Solid lines represent mean cross-validation performance, while dashed lines denote test performance. Error bars indicate the standard deviation across cross-validation folds. Model performance generally improves as additional features are incorporated and stabilizes when larger feature sets are used.}
	\label{fig:perf_vs_k}
\end{figure}

\end{document}